\documentclass[a4paper,fleqn,usenatbib]{mnras}
\usepackage[T1]{fontenc}
\usepackage{ae,aecompl}
\usepackage{graphicx} 
\usepackage{amsmath}
\usepackage{siunitx}
\usepackage{amssymb} 
\usepackage{caption}
\usepackage{subcaption}
\title[A VLBA-uGMRT study of DPAGN with X-shaped radio morphology]{A VLBA-uGMRT search for candidate binary black holes: Study of six X-shaped radio galaxies with double-peaked emission lines}
\author[Sebastian et al.]{Biny Sebastian,$^{1}$\thanks{E-mail: Biny.Sebastian@umanitoba.ca}
Anderson Caproni,$^{2}$
Preeti Kharb,$^{3}$
Nayana A.J.,$^{4,5}$
Arshi Ali,$^{6}$
\newauthor{K., Rubinur,$^{7}$
Christopher P. O'Dea,$^{1}$ Stefi Baum,$^{1}$ Sumana Nandi$^{7}$}
\\
$^{1}$Department of Physics and Astronomy, University of Manitoba, Winnipeg, MB R3T 2N2, Canada \\
$^{2}$N\'ucleo de Astrof\'\i sica, Universidade Cidade de S\~ao Paulo R. Galv\~ao Bueno 868, Liberdade, S\~ao Paulo, SP, 01506-000, Brazil \\
$^{3}$National Centre for Radio Astrophysics (NCRA) - Tata Institute of Fundamental Research (TIFR), S. P. Pune University Campus,\\ Post Bag 3, Ganeshkhind, Pune 411007, India\\
$^{4}$Department of Astronomy, University of California, Berkeley, CA 94720-3411, USA\\
$^{5}$Indian Institute of Astrophysics, Block II, Koramangala, Bangalore 560 034, India\\
$^{6}$Department of Physics, Savitribai Phule Pune University, Pune 411007, India\\
$^{7}$Institute of Theoretical Astrophysics, University of Oslo, P.O box 1029 Blindern, 0315 OSLO, Norway\\
$^{8}$Manipal Centre for Natural Sciences, Centre of Excellence, Manipal Academy of Higher Education, Manipal, \\ Karnataka - 576104, India}

\date{Accepted XXX. Received YYY; in original form ZZZ}
\pubyear{2023}

\begin{document}
\label{firstpage}
\pagerange{\pageref{firstpage}--\pageref{lastpage}}
\maketitle

\begin{abstract}
Identifying methods to discover dual AGN has proven to be challenging. Several indirect tracers have been explored in the literature, including X/S-shaped radio morphologies and double-peaked (DP) emission lines in the optical spectra. However, the detection rates of confirmed dual AGN candidates from the individual methods remain extremely small. We search for binary black holes in a sample of six sources that exhibit both X-shaped radio morphology and DP  emission lines using the VLBA. Three out of the six sources show dual VLBA compact components, making them strong candidates for binary black hole sources. In addition, we present deep uGMRT images revealing the exquisite details of the X-shaped wings in three sources. We present a detailed precession modeling analysis of these sources. The BH separations estimated from the simplistic geodetic precession model are incompatible with those estimated from emission line offsets and the VLBA separations. However, precession induced by a noncoplanar secondary black hole is a feasible mechanism for explaining the observed X-shaped radio morphologies and the black hole separations estimated from other methods. The black hole separations estimated from the double-peaked emission lines agree well with the VLBA compact component separations. Future multi-frequency VLBA observations will be critical in ruling out or confirming the binary black hole scenario in the three galaxies with dual component detections.
\end{abstract}

\section{Introduction}
According to the hierarchical galaxy formation scenario, galaxy mergers play a significant role in galaxy dynamics and evolution. It is easy to visually identify systems that are at the early stages and have dual optical cores separated by tens of kpc \citep{Foreman2009, Myers2007, Hennawi2006, Darg2010a}. However, the dynamics of the later stages of galaxy mergers are hard to study because of several observational limitations. As the cores' separation becomes less, resolving them individually becomes harder. The obscuration levels due to dust are quite high in the optical cores of merging systems compared to those in isolated galaxies \citep{Liu2013, Kocevski2015}, making optical identification more difficult. \cite{Barnes1991} have suggested that mergers result in the funneling of gas to the central supermassive black holes (SMBHs). This gas inflow initiates the activity of the active galactic nucleus \citep[AGN;][]{Hopkins2005, Hopkins2008} and could also lead to a starburst near the core. \cite{Comerford2015} have classified SMBH binaries separated by a few kpcs during galaxy mergers as dual AGN (DAGN) if both the black holes are active and as ``offset AGN'' if only one of them is active. Efforts to identify DAGN systems have often seen little success, and so far, there is only a handful of confirmed DAGN reported in the literature \citep[e.g.,][]{Deane2014, Balmaverde2018, Husemann2020}. 

Indirect indicators of binary black holes (BBHs) at the center of galaxies are double-peaked (DP) emission-lines \citep{Blecha2013,DoanAnh2020,Kharb2020}, periodicity in optical \citep{Bon2012, Pihajoki2016, Graham2015} or radio light curves \citep{ONeill2022} and X- or Z-shaped radio sources \citep{zhang2007, Kharb2017}. Systematic searches to identify DP emission lines in galaxies were carried out by several authors \citep{Liu2010a, Smith2010, Ge2012, Fu2012, Kharb2021}. DP emission lines may occur due to several reasons, including separate NLR around two massive black holes at the center, peculiar gas kinematics in a disk-like narrow line region \citep{Fu2009} and bipolar outflows, which are driven either by AGN radiation or starbursts \citep{Crenshaw2010} or jet-cloud interaction \citep{Stockton2007, Rosario2010, Kharb2017b, Kharb2019}. While the first two reasons are merger-related, the bipolar outflows are not induced by recent merger activity. Hence, the mere presence of these DP emission lines in a galaxy does not guarantee the presence of DAGN. Observations at wavelengths where the nuclei are unobscured, like the X-rays or high-resolution radio observations, are needed to confirm DAGN \citep{Komossa2003, Bianchi2008, Hudson2006}. There is only one parsec-scale DAGN detected using Very Long Baseline Interferometric observations \citep[VLBI;][]{Rodriguez2006}. It is worth noting that follow-up high-resolution imaging and spatially resolved spectroscopy of the double-peaked AGN (DPAGN) samples by \cite{Shen2011} and \cite{Fu2011} did not find DAGN in a majority of their sources.
 
Previous radio studies of DPAGN have mostly relied on high-resolution imaging studies to resolve the dual radio cores \citep{Tingay2011,muller2015}. The large-scale radio morphologies for a sample of DPAGN have not been explored in detail so far. There are several individual source studies where DPAGN are also hosted by X/S shaped radio galaxies \citep{Wang2009, Rubinur2017, zhang2007, Nandi2021}. Three major formation mechanisms for X/S-shaped radio sources have been proposed in the literature. These include realignment of the jets \citep{Ekers1978, Klein1995, Rees1978}, backflow diversion \citep{Leahy1984} and twin AGN \citep{Lal2005, Lal2007, Lal2019}. The realignment of jets is believed to occur either due to the precession of the jet or a spin flip of the SMBH powering the jets, which in turn is induced by coalescence of a BBH \citep{Dennett2002, Zier2005, Merritt2002, Rottmann2001}. 

The precession of jets from BBHs on galactic scales is known to exist. This precession occurs due to the gravitational influence of the second member in the binary. Similarly, in the case of extragalactic jets, the precession has been attributed to the presence of a binary black hole pair at the center of the galaxy \citep{Begelman1980}. These black holes are remnants of two merged galaxies, each initially having a supermassive black hole at the center. Yet another explanation for jets' precession is the warping of accretion disks \citep{Liu2004}.

BBH scenario forms one among several possible explanations for the appearance of both DP emission lines and cross-symmetric morphology. While isolated cases of DPAGN hosting X or S-shaped radio galaxies were reported in the past, a systematic study of a comprehensive sample has not yet been done. If mergers or BBHs lead to both the X-shaped radio morphology and the dual AGN formed from the merger leads to the DP emission features, we expect that a sample of DPAGN with X/S-shaped radio morphology should improve the probability of the detection of dual AGN.

 The expected black hole separations for a sample of sources selected on the basis of these two criteria are only a few parsecs. Hence, VLBA proves to be the ideal telescope to zoom into the centers of such merger systems. \cite{Begelman1980} have shown that a dual AGN would remain at $>$ 1 kpc separation before getting closer for at least 100 Myrs. Afterward, the timescales of evolution to $\sim$1~pc separation via the scattering of stars in the nucleus could be anywhere between 10$^8-10^9$ years (the loss-cone problem). 
Such large timescales mean that we need not see merger signatures surrounding the host galaxy, such as tidal features, as they are usually detectable only up to 200$\sim$400 Myrs \citep{Lotz2010}. Hence, detecting and confirming BBHs at $\sim$pc scale separations is extremely hard. Figure~1 from \cite{Chen2022} illustrate this rarity of double quasars reported in the literature below a separation of 1~kpc. However, the double peaks in optical spectra arising from rotating disks due to major mergers arise around 1~Gyr after the final coalescence \citep{Maschmann2023} and could be a better tracer of pc-scale binaries.


In this paper, we have studied a sample of six galaxies possessing both a cross-symmetric radio morphology and double-peaked emission lines in their optical spectra using the Very Long Baseline Array (VLBA) and upgraded Giant Metrewave Radio Observations (uGMRT). The paper tests whether our sample selection criteria are robust for identifying DAGN. The paper is organized as follows. We describe our sample selection in Section~\ref{sample} and provide the observational details and the data analysis in Section~\ref{sec:obs}. In Section~\ref{sec:res}, we present the results of various analyses on our sample sources. We discuss the implications of our results in Section~\ref{sec:disc} and finally summarize our conclusions in Section~\ref{sec:summ}.

\section{Sample selection}
\label{sample}
\cite{Ge2012} had searched for DP emission lines in the SDSS DR7 spectra \citep{Abazajian2009} of a sample of 920,000 galaxies. They had visually inspected a sub-sample of 42,927 candidate DP emission line galaxies, which possessed two Gaussian components in their emission spectra. They presented a final sample of 3030 DP narrow-emission line galaxies, ensuring the presence of a clear trough between two narrow emission lines. We cross-matched the location of the optical host galaxies from this complete sample of 3030 DP narrow emission-lines galaxies listed in \cite{Ge2012} with the VLA FIRST survey catalog \citep{Becker1995} images within a distance of one arcmin. We then visually inspected the radio images of the resulting 643 cross-matches using VLA FIRST, TGSS \citep{Intema2017}, and NVSS surveys and found that 30 sources showed extended radio emission. Out of these 30, four showed cross-symmetric morphology. The large-scale radio structure in J1328+27 has already been studied by \cite{Nandi2017}. We obtained uGMRT observations for the three remaining sources, viz, B2\,1059+29, 4C+32.25 and 4C\,61.23, at Band~3 ($250-300$~MHz) and Band~5 ($1000-1450$~MHz). 
 
In addition to the above sources, \cite{Wang2009} had reported that 3C\,223.1, and J1430+54 show DP narrow emission lines and X-shaped radio morphology. 
Hence, we observed all these six potential binary black hole candidates with the VLBA to uncover dual AGN. 
We obtained 5~GHz VLBA observations for the four sources, including the three sources that did not have any phase-referenced VLBA archival data. We observed the two sources that already had archival VLBA data, namely, 4C +32.25 and 4C +61.23 at 15 GHz. 
A detailed analysis of the VLBA observations of J1328+2752 has already been presented by \cite{Nandi2021}. We summarize the results of the remaining sources in the current paper.

\section{Observations and data analysis}
\label{sec:obs}
\subsection{VLBA Observations: 5 and 15 GHz}
Data were acquired with nine antennas of the VLBA in a phase-referencing mode on 11 August 2018 at 4.85~GHz and 13 August 2018 at 15.13~GHz (Project IDs: BS267B and BS267A, respectively). Pie Town (PT) did not participate in the experiment. The total bandwidth of the data was 256~MHz. The calibrators DA193 and J1800+3848 were used as the fringe finders for the experiment. The phase calibrators used were J0819+3226, J1128+5925 at 15 GHz, and J1333+2725, J0934+3926, J1103+3014, J1437+5112 at 5 GHz. The data were reduced using the VLBARUN data-reduction pipeline with appropriate parameters inside AIPS, with Los Alamos (LA) as the reference antenna. Five sources, viz. 4C\,32.25 (at 15~GHz) and J1328+2752, 3C\,223.1, J1430+5217 and B2\,1059+29 (at 5~GHz) were detected. 

\subsection{uGMRT Observations: Band 3 and Band 5}
\label{obs}

Details of the observations of three sources that were observed using the uGMRT at Band-3 and Band-5 are provided in Table~\ref{tab:gmrt}.
The data analysis used the uGMRT data analysis pipeline {\tt aipsscriptwriter}\footnote{https://github.com/binysebastian/aipsscriptwriter}. It uses both {\tt AIPS}and {\tt CASA} tasks to carry out the initial editing and flagging of bad data.
The pipeline calibrates the data using standard procedures in {\tt AIPS}. 
The target source is split out from this dataset while averaging several channels. The initial few rounds of imaging and phase self-calibration are done in {\tt AIPS}, after which the final imaging is done using the MT-MFS algorithm \citep{Rau2011} that is available in {\tt CASA}.
The GMRT images of the three galaxies in bands 3 and 5 are shown in Figures~\ref{gmrt4c32.25} and \ref{gmrtb2and4c61}.

\subsection{SDSS Optical Spectral Modeling}
\label{sec:specmodel}
 The modeling of the optical spectra for all sources except J1430+5217 was already carried out and presented by \cite{Ge2012} and \cite{Wang2009}. While \cite{Wang2009} (see Section~4.3) specifies the presence of double-peaked emission lines in the SDSS spectra of J1430+5217, they do not provide the emission line modeling results for this galaxy. Hence, in this paper, we perform spectral modeling and the double Gaussian fitting for the emission lines in J1430+5217.
We used the SDSS spectra with wavelength coverage from $\sim$ 4000~\AA - 9000~\AA. 

To perform the spectral fitting for 
J1430+5217, we have used the open-source Python code Bayesian AGN Decomposition Analysis for SDSS Spectra pipeline {\tt BADASS3}\footnote{https://github.com/remingtonsexton/BADASS3} \citep{Sexton2021}. 
{\tt BADASS3} performs modeling for various components (e.g., power-law continuum, stellar line-of-sight velocity distribution, emission lines, and potential outflows). 
For the analysis presented in this paper, we only require the emission line parameters of [O\,III] and H$\beta$ line. Hence, we restricted the wavelength range to 4800~\AA to 5050~\AA.
We used the empirical stellar template Indo-US \citep{Valdes2004} to fit the continuum emission. The {\tt BADASS3} pipeline iteratively refines the estimated parameters and components until the best fit to the original data is achieved. 
We use the redshift of the source to determine the rest frame velocity of the galaxy. For J\,1430+5217, the peak of the absorption lines is consistent with the rest frame velocity within the error bar.
For the rest of the sources, \cite{Ge2012} use absorption lines to estimate the systemic velocity.

{\tt BADASS3} treats multiple line components as outflows, and hence, we model the emission lines as a combination of narrow-line and outflow components. We assume Gaussian profiles for both these components. Figure~\ref{fig:J1430+5217} shows the raw spectra, the model, and individual components, including the stellar continuum and the multiple Gaussian narrow emission lines. Even though we modeled the emission lines using a narrow + outflowing component, neither of these components seems to be centered around the systemic velocity of the host galaxy, pointing to the existence of double-peaked emission lines. We have tabulated the results of this emission line fitting required for the analysis in this paper in Table~\ref{tab:BH-mass-estimates-nelson}.


\begin{table*}
\begin{center}
\caption{Sources and observed properties}
\tabcolsep=0.11cm
\begin{tabular}{cccclccc}
\hline \hline
{VLBA observations} \\ \hline
Source  &Alternate& Redshift&  Observation & $\nu_{\rm cen}$ & Beam, PA & Image peak flux & Image r.m.s \\
  name   &name   & ($z$)&  date & (GHz) & (mas, $\degr$) & density (mJy) & (mJy~beam$^{-1}$) \\
\hline
4C\,+32.25	& J0831+3219    & 0.05121 & 2018 Aug 13	& 15.26	& 1.07$\times$0.49, $-$4.17	& 0.6	& 0.06  \\
3C\,223.1	& J0941+3944    & 0.10747 & 2018 Aug 11	& 4.98	& 3.46$\times$1.62, 15.56	& 2.9	& 0.02  \\
B2\,1059+29	& J1102+2907    & 0.10598 & 2018 Aug 11	& 4.98	& 3.4 $\times$1.53,  1.35	& 0.1	    & 0.02  \\
4C\,+61.23	& J1137+6120    & 0.11110 & 2018 Aug 13	& 15.26	& 1.06$\times$0.52, 29.18	& 0.4	& 0.06  \\
J1328+27	& J1328+2752     & 0.09    & 2018 Aug 11	& 4.98	& 3.44$\times$1.52, $-$0.34	& 0.3	    & 0.02  \\
J1430+52	& J1430+5217     & 0.3671  & 2018 Aug 11	& 4.98	& 3.56$\times$1.51, $-$15.4 & 4.9       & 0.02  \\
\label{tab:vlba}
\end{tabular}
\begin{tabular}{cclccc} \hline \hline
{uGMRT observations} \\ \hline
Source  & Observation & $\nu_{\rm cen}$ & Beam, PA & Image peak flux & Image r.m.s \\
 name       & date & (MHz) & (arcsec, $\degr$) & density (mJy~beam$^{-1}$) & (mJy~beam$^{-1}$) \\

\hline
4C +32.25 &     2017 Nov 04  & Band-3 (250-500)   & $12.2\times7.7$, 80.42 & 123.9& 0.08 \\  
          &     2017 Oct 28  & Band-5 (1000-1450) &  $6.2\times3.9$, 64.6 & 26.5& 0.15 \\      
B2 1059+29&     2017 Oct 27  & Band-3 (250-500)   &$12.3\times8.5$, 49.03&85.5	 & 0.76 \\ 
          &     2017 Oct 28  & Band-5 (1000-1450) & $5.1\times2.6$, 76.6 &21.2	 & 0.17 \\ 
4C +61.23 &     2017 Nov 04  & Band-3 (250-500)   & $12.2\times8.1$, 49.03  &146.3 & 0.19 \\ 
          &     2017 Oct 28  & Band-5 (1000-1450) &  $4.0\times3.4$, $-48.41$ &32.3	 & 0.07 \\ 
         \hline
\label{tab:gmrt}
\end{tabular}
\end{center}
\end{table*}

\section{Results}
\label{sec:res}
We describe the results of the uGMRT and VLBA observations below. We present the results for individual sources first, followed by a global analysis.

\subsection{Radio Morphology with the uGMRT}
\subsubsection{4C\,32.25} 

Our total intensity maps of 4C\,32.25 at 400 MHz and 1250 MHz are presented in Figure ~\ref{gmrt4c32.25}. The high resolution offered by uGMRT at low frequencies and enhanced sensitivity with the bandwidth upgrade enabled us to map the detailed structure of the diffuse emission in both the primary and the secondary lobes. We can distinctly spot the core, the jet towards the east, and the counterjet towards the western side, as inferred from the brightness ratio. 
The jets are not perfectly linear and possess a significant curvature. Our maps also reveal the extended nature of both hotspots, a feature not witnessed commonly in radio galaxies. The eastern hotspot has a linear extension towards the northwest, aligned almost at right angles to the jet.
The nonlinearity of the jets, the extension of the hotspots, and the presence of secondary lobes in the source vouch for a dynamic model for the X-shape in the source. 

\begin{figure*}
\includegraphics[height=8cm,trim = 0 0 0 0 ]{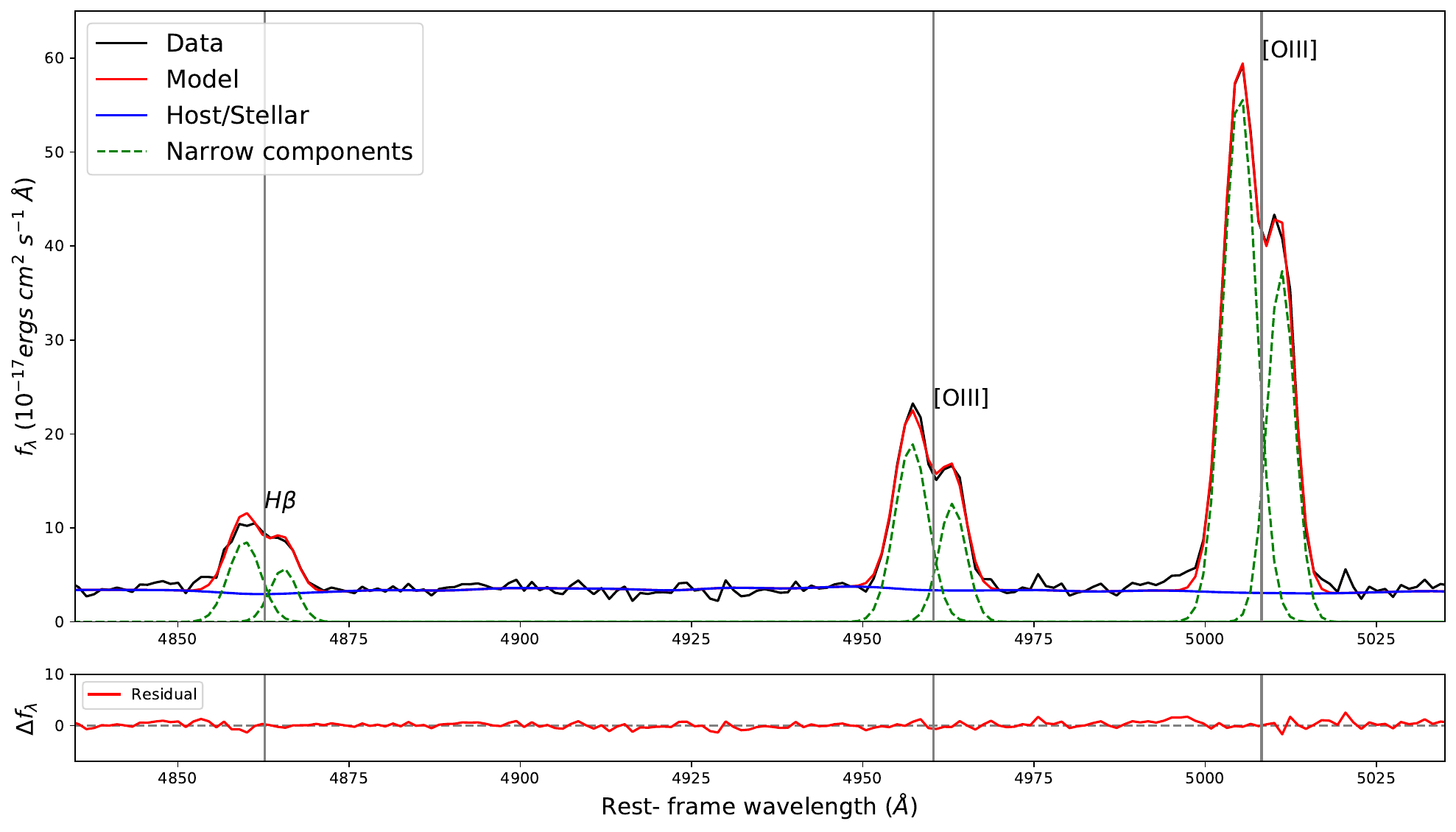}
\caption
{\small Spectral fitting of J1430+5217 for H$\beta$+[O\,{\small{III}}] complex lines
using the {\tt BADASS} \citep{Sexton2021}. The emission lines have been fitted using double Gaussian profiles. The solid red line represents the overall model obtained by fitting, the solid blue line represents the combined continuum level, which includes both the host-galaxy and AGN continuum contributions, and the dashed green lines correspond to the recognized narrow emission lines considered in the fitting process. The grey solid line corresponds to the rest frame wavelength of the emission lines.}
\label{fig:J1430+5217}
\end{figure*}

\begin{figure*}
\begin{center}
\includegraphics[width=8.1cm,trim=40 100 0 120]{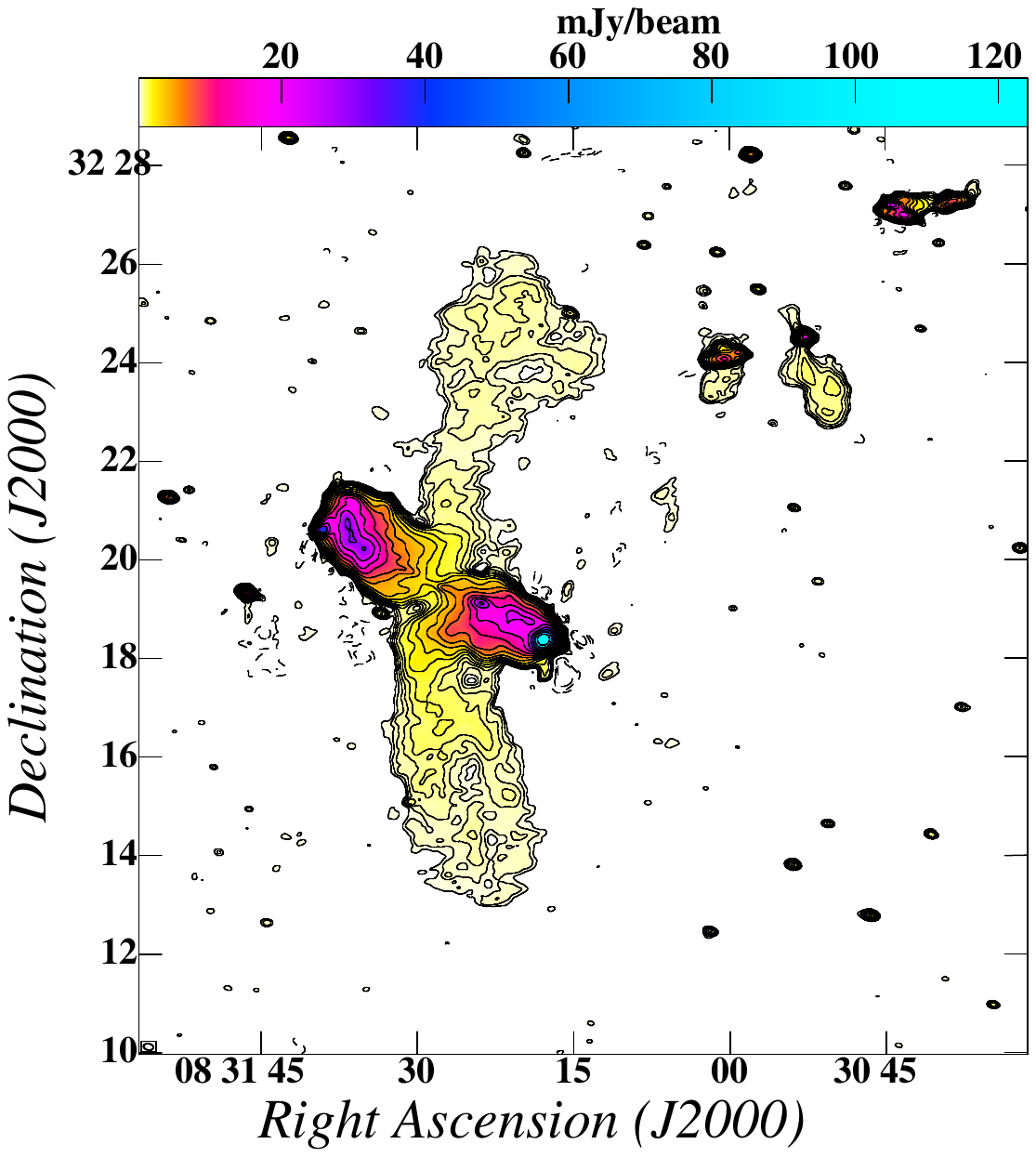}
\includegraphics[width=9.5cm,trim=10 130 0 120]{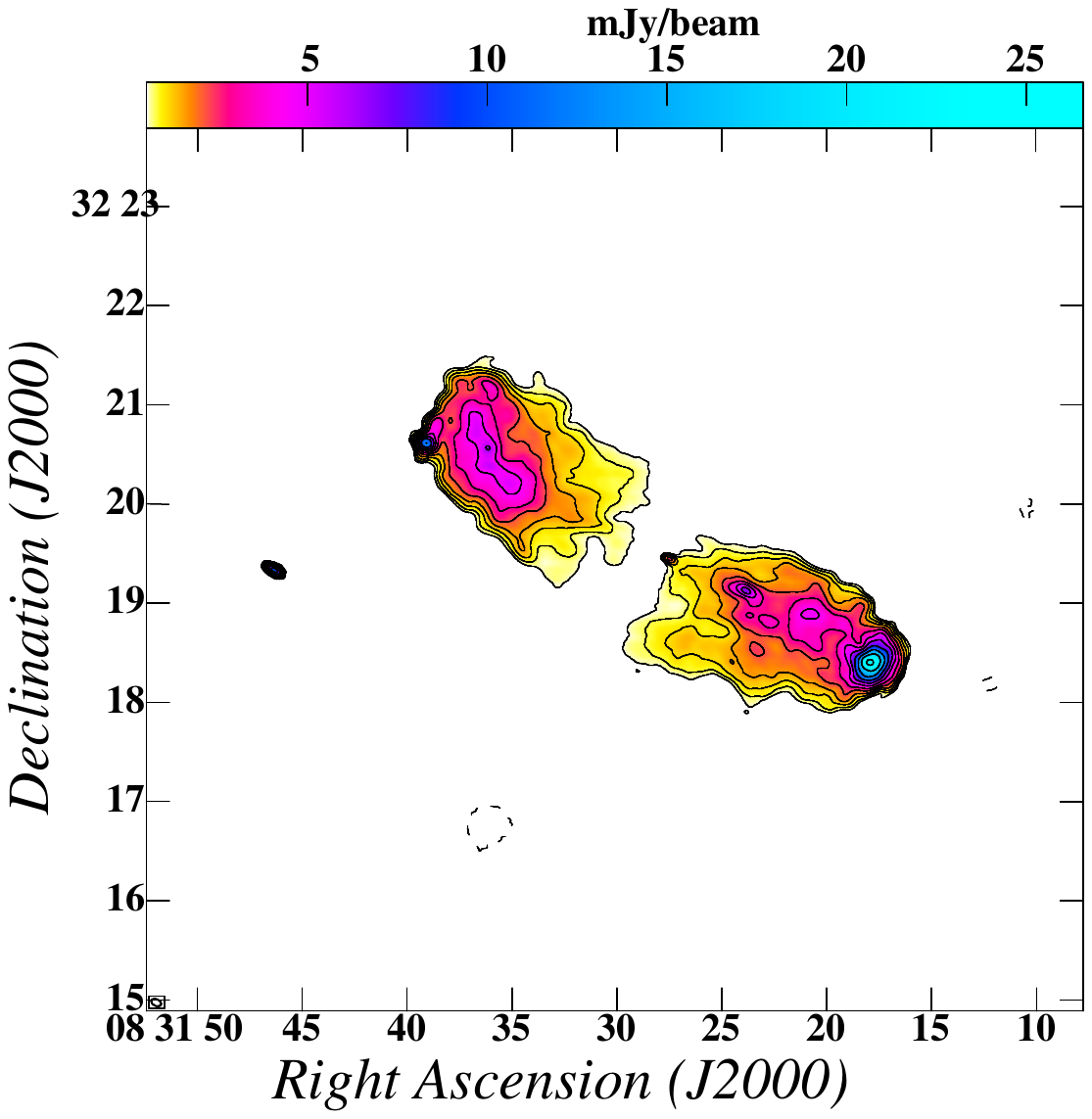}
\caption
{\small \textit{Left:} uGMRT image of 4C+32.25 at 400 MHz.  Contour levels are 0.23~mJy$\times$(-2, -1.4, -1, 1, 1.4, ...). \textit{Right:} uGMRT image of 4C+32.25 at 1250~MHz. Contour levels are 0.67~mJy$\times$(-2, -1.4, -1, 1, 1.4, ...). The image resolution and rms are provided in Table~\ref{tab:gmrt}. The beam FWHM is shown at the bottom left corner of the images.}
\label{gmrt4c32.25}
\end{center}
\end{figure*}

\begin{figure*}
    \includegraphics[height=9.5cm,trim = 10 60 0 100]{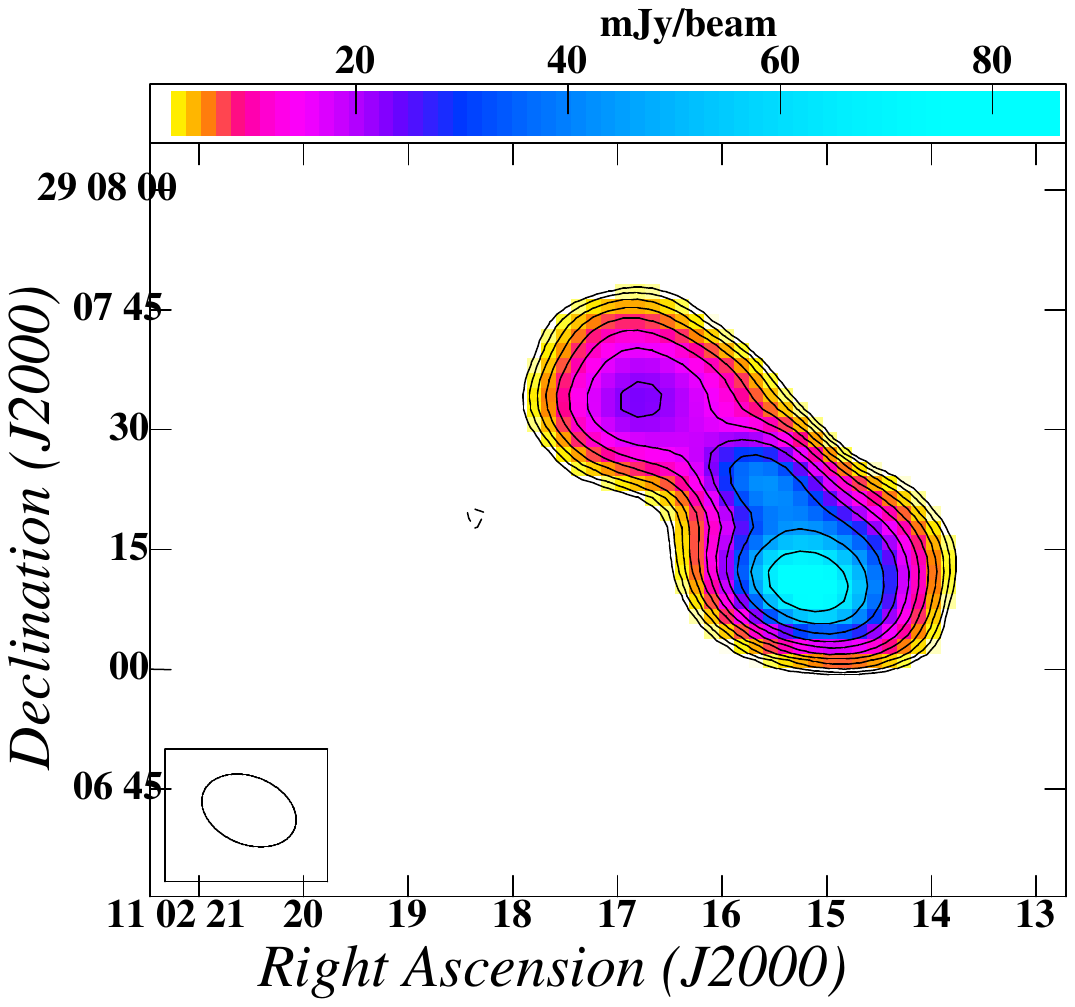}
    \includegraphics[height=8.9cm,trim = 15 30 0 100]{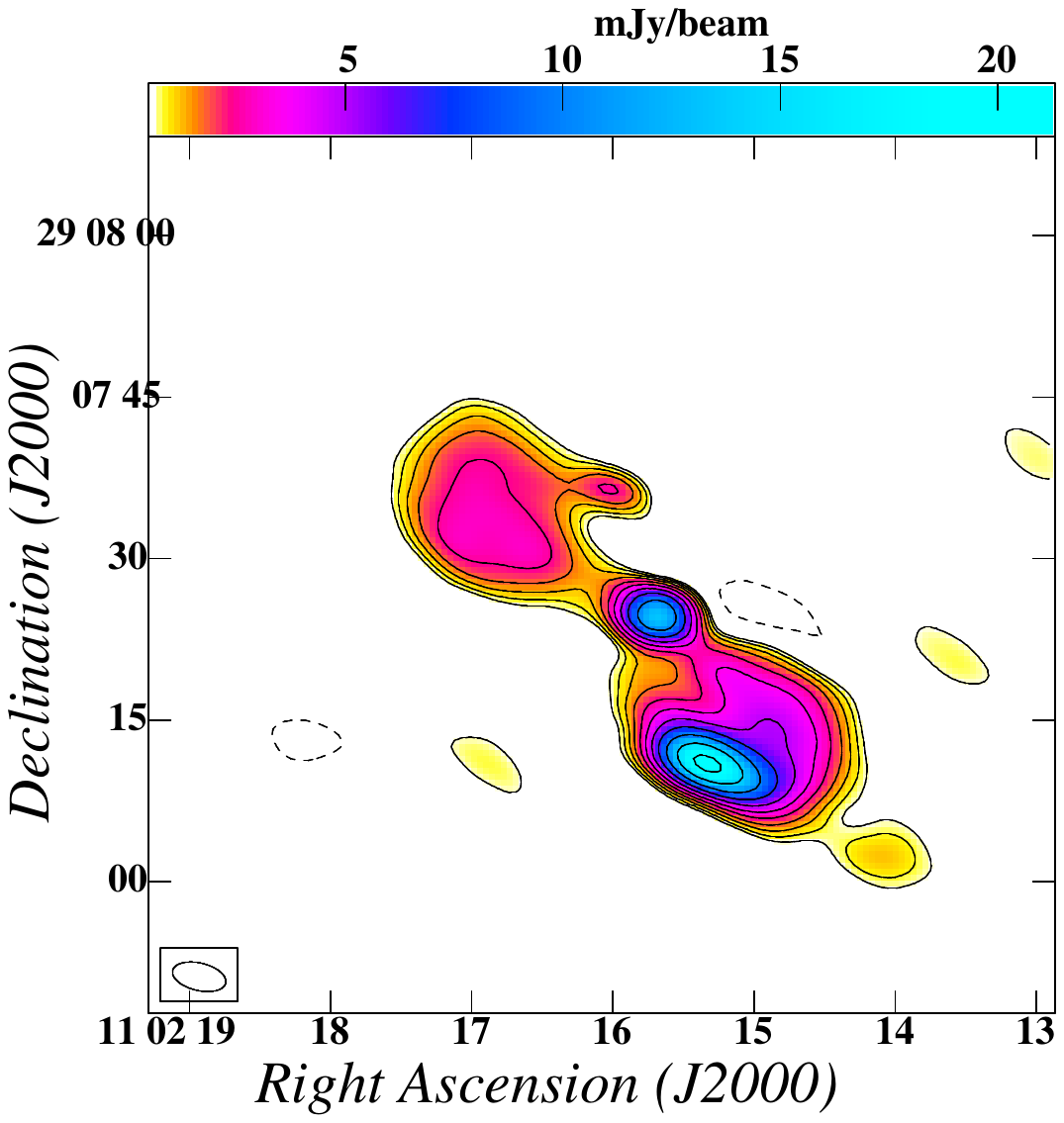}
    \includegraphics[height=7.2cm,trim=0 100 0 200]{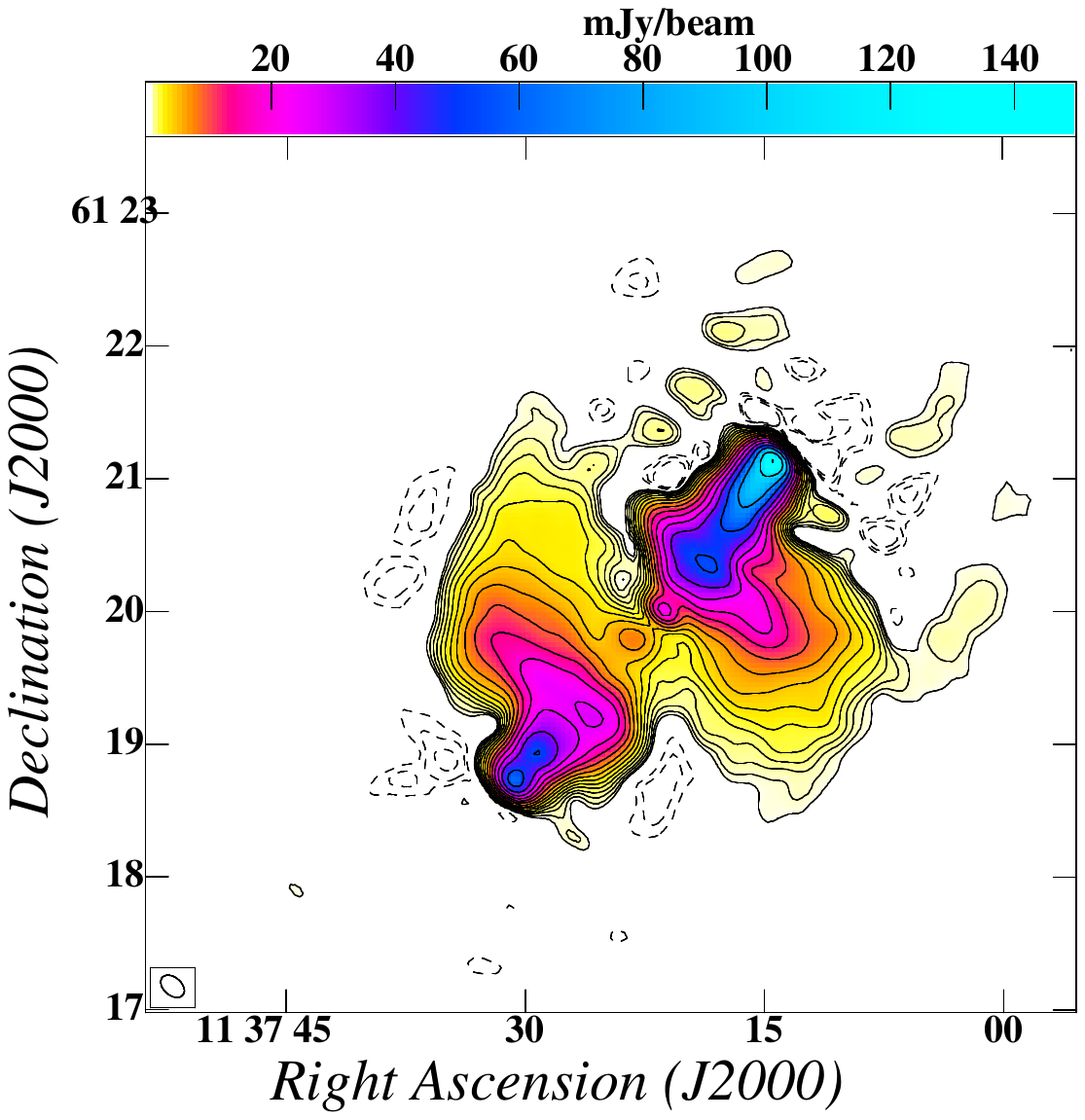}
    \includegraphics[height=5.7cm,trim=0 0 0 170]{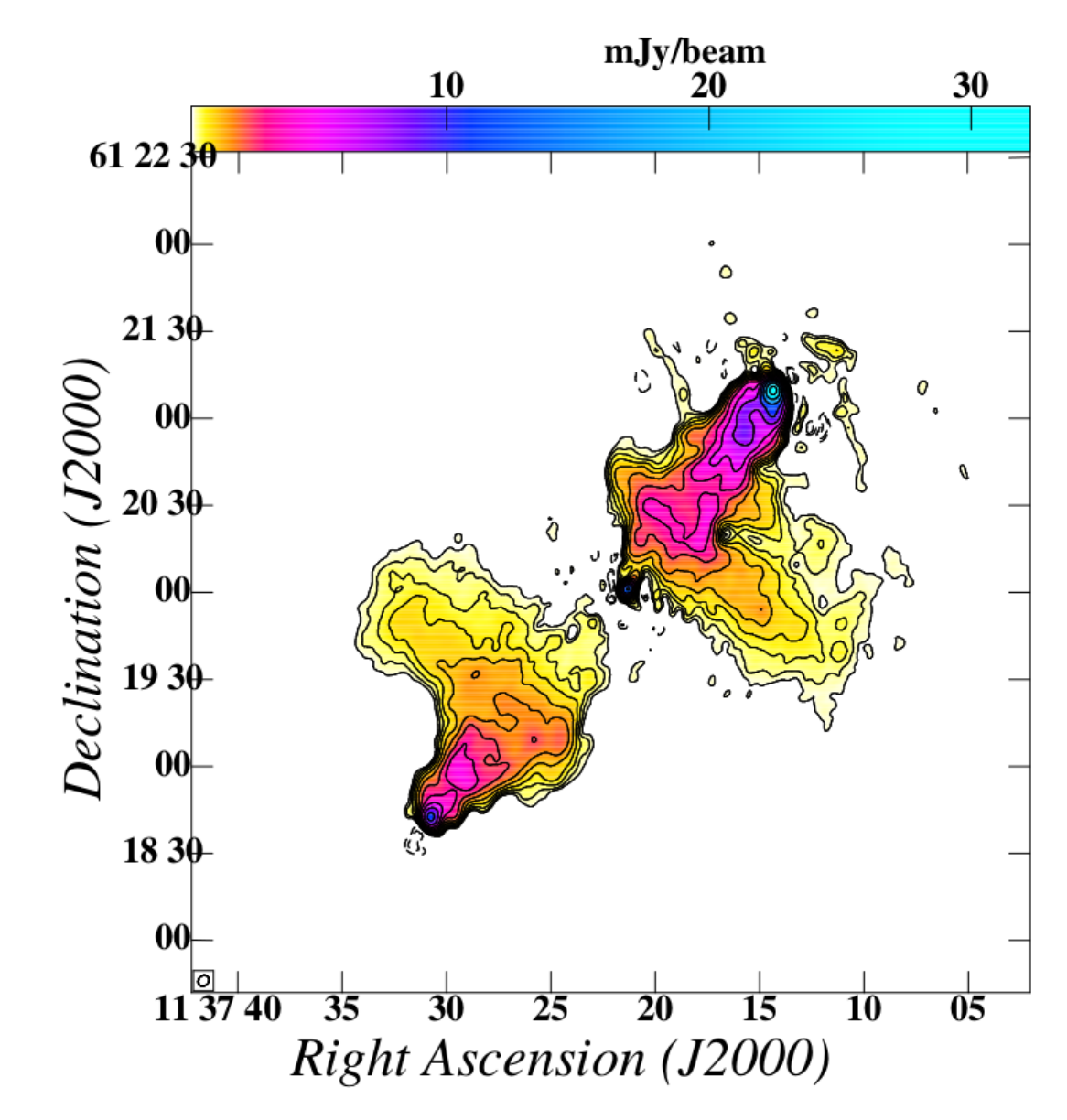}
    \caption{\small \textit{Top~left:} uGMRT images of B2\,1059+29 at 400~MHz. Contour levels are 2~mJy$\times$(-2, -1.4, -1, 1, 1.4, ...). \textit{Top right:} uGMRT image of B2\,1059+29 at 1250~MHz. Contour levels shown are 0.6~mJy$\times$(-2, -1.4, -1, 1, 1.4, ...). \textit{Bottom left:} uGMRT image of 4C+61.23 at 400~MHz. Contour levels are 0.41~mJy$\times$(-2, -1.4, -1, 1, 1.4, ...). \textit{Bottom right:} uGMRT image of 4C+61.23 at 1250~MHz. Contour levels are 0.31~mJy$\times$(-2, -1.4, -1, 1, 1.4, ...). The image resolution and rms are provided in Table~\ref{tab:gmrt}. The beam FWHM is shown at the bottom left corner of the images.}
    \label{gmrtb2and4c61}
\end{figure*}

\begin{figure*}
\includegraphics[width=8cm,trim=0 0 0 70]{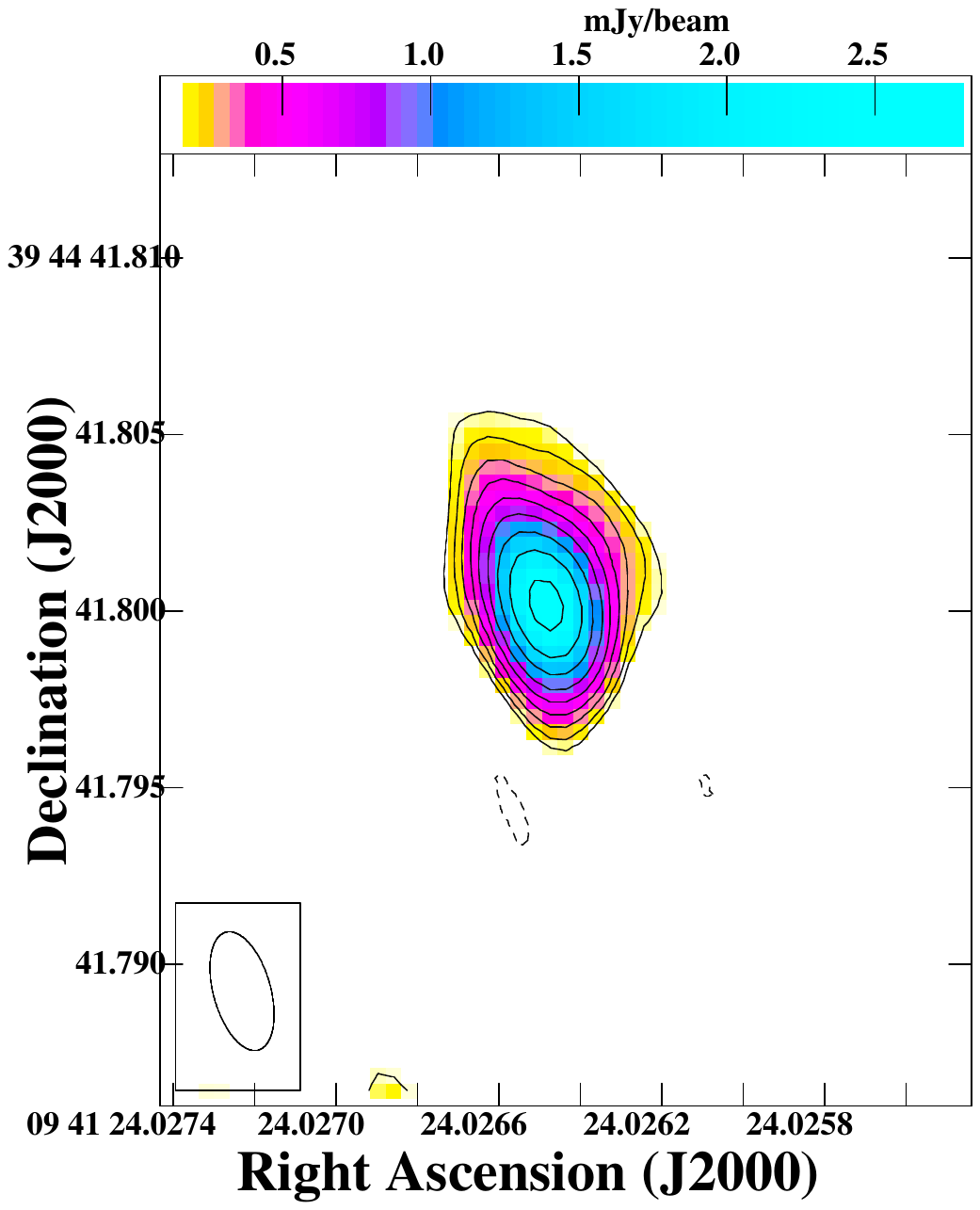}
\includegraphics[width=8.4cm,trim=0 20 0 70]{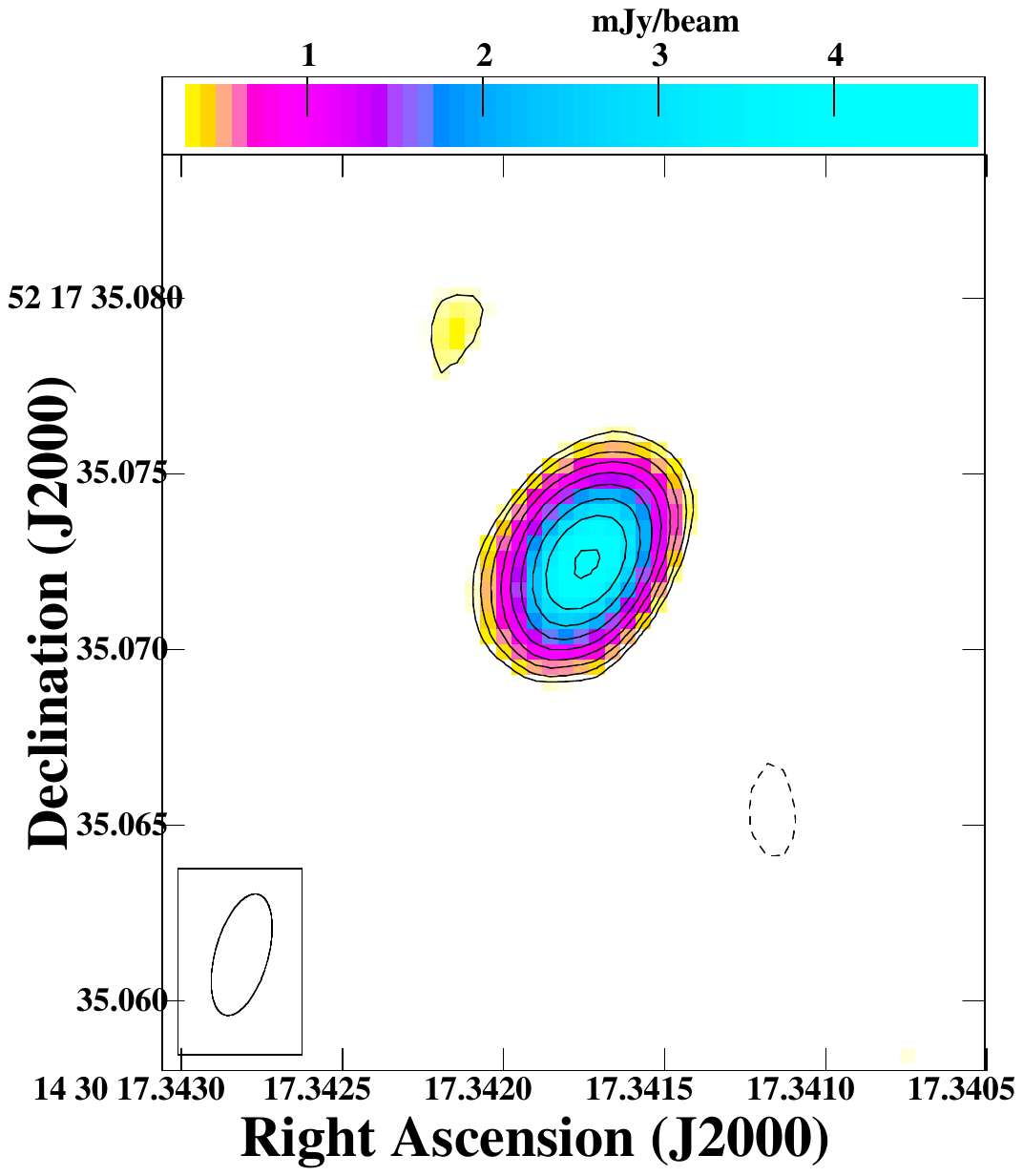}
\includegraphics[width=8.3cm,trim=0 50 0 130]{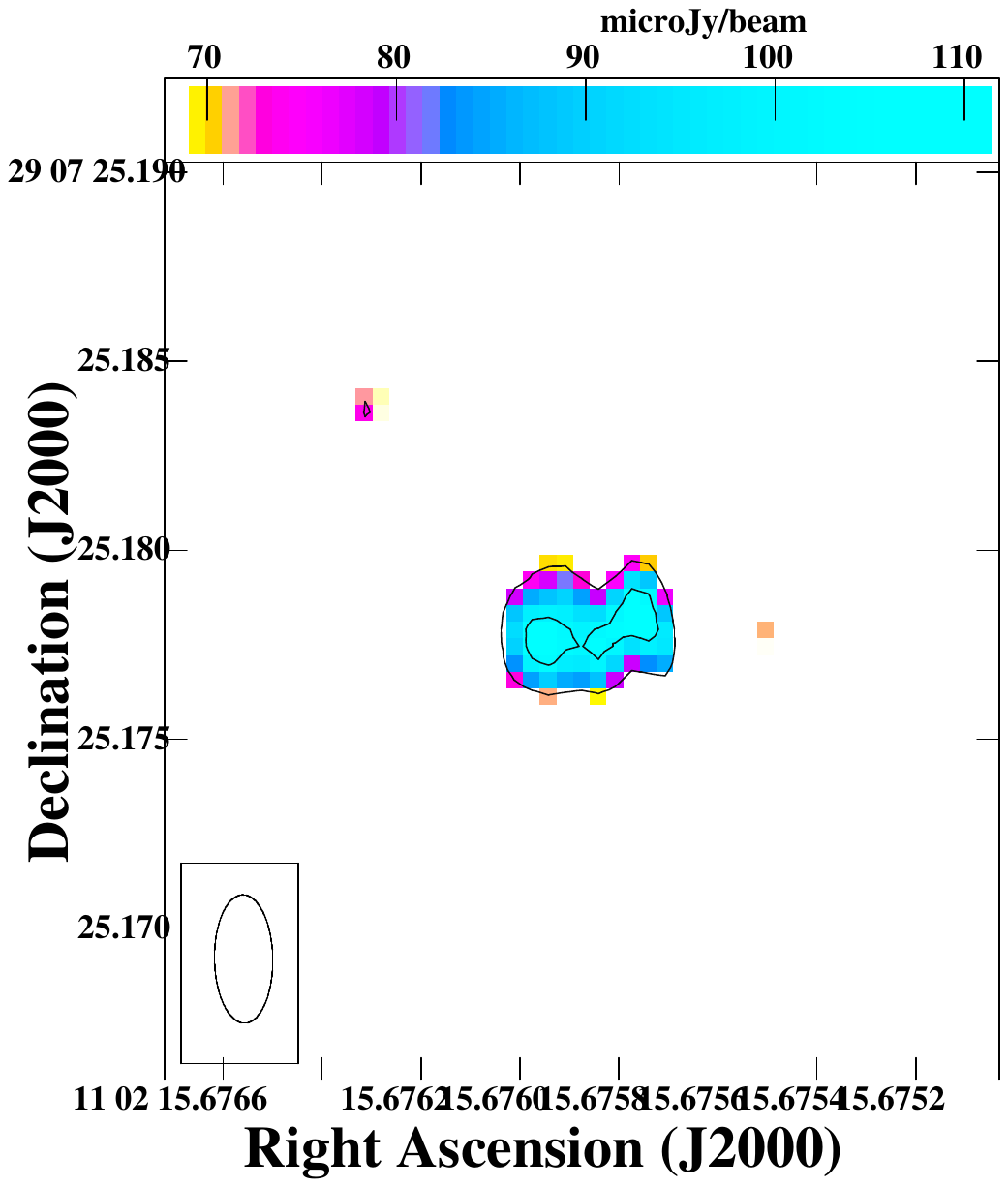}
\includegraphics[width=8.6cm,trim=20 80 0 130]{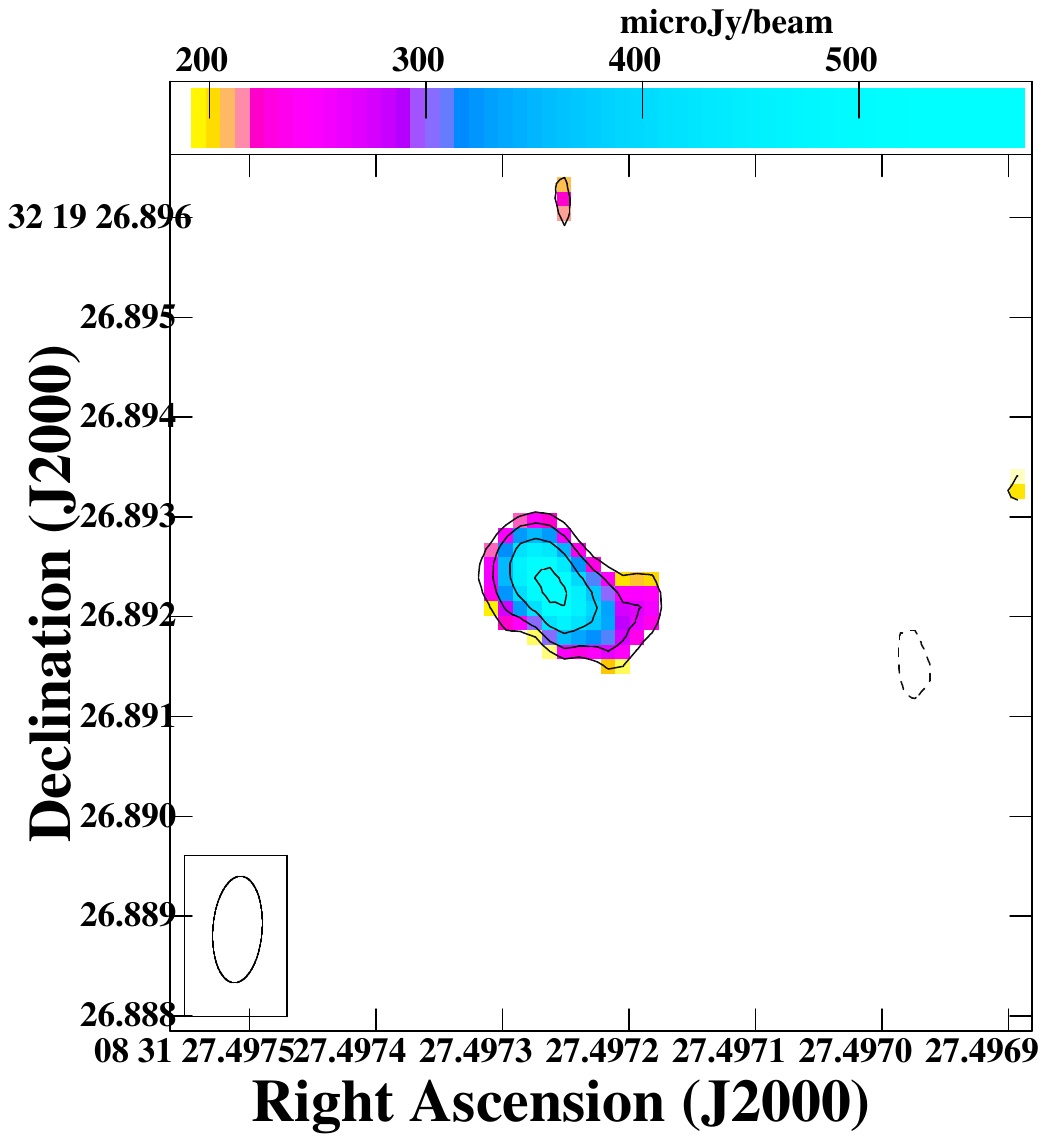}
\caption {\small \textit{Top left}: 5~GHz VLBA image of 3C223.1. 
\textit{Top right}: 5~GHz VLBA image of J1430+52. 
\textit{Bottom left}: 5~GHz  VLBA image of B2\,1059+059. 
\textit{Bottom right}: 15~GHz VLBA image of 4C32.25. The image rms and resolution are provided in Table~\ref{tab:vlba}. The beam FWHM is shown at the bottom left corner of the images.}
\label{vlba}
\end{figure*}

\subsubsection{B2\,1059+29} 
B2\,1059+29 is at a redshift of $z$ = 0.106. The SDSS spectrum of the source shows double-peaked [O {\small III}] lines. The Legacy Survey image of  B2 1059+29 shows tidal features suggesting a past merger event. 
The source appears to possess a rotational symmetry in the low-resolution images. The detailed structure of the source is conspicuous in our higher-resolution images. The jet direction is seen to be changing continuously throughout the length of the jet. Very close to the core, there is an elongation of the jet in the north-south direction. This is consistent with the jet seen using the VLA high-resolution image \citep{Rubinur2019}, which is aligned along the north-south direction. The jet seems to change its alignment on larger scales, and the primary jet, as seen in our L-band images, is aligned along the NE-SW direction. The jet seems to change its direction resulting in the formation of an S-shaped structure. 

\subsubsection{4C\,61.23} 


4C+61.23 was classified as an FR-II radio galaxy using the 5 GHz VLA image \citep{Lara2001}. The GMRT maps at both bands show clear `X' shaped morphology of this source. The primary and secondary lobes are visible. From visual inspection, it looks like there is a continuity in the emission from the brighter lobes to the fainter ones.

\subsection{Dual Compact Components in VLBA: Candidate BBHs}
\label{sec:dualcompactvlba}
Three of the six sources in our sample show evidence for dual radio compact components in at least one frequency image with the VLBA. All three sources, therefore, are possible BBH candidates that need to be confirmed with follow-up observations. We do not detect a second component in the rest of the three sources, namely, 3C\,223.1, J1430+52, and 4C\,32.25. However, 3C\,223.1 and 4C\,32.35 show extended pc scale morphology, which might be due to the core-jet system. We discuss the VLBA morphology of the three sources exhibiting dual-components below.

\subsubsection{B2\,1059+29}
This source shows two close, faint, compact components in the VLBA image (see Figure~\ref{vlba}). The central point-source spectrum estimated using our uGMRT images at 400~MHz and 1250~MHz is steep ($\alpha\sim-0.8$), diminishing the likelihood of the BBH scenario. Since the central point-source spectrum is steep, the extended structure seen in the VLBA image might likely be jet-related, and the source morphology closely resembles a compact symmetric object (CSO). If accurate, the misalignment of the inner double with the outer lobe will provide clues to distinguish between the backflow and realignment models for X-shaped sources. However, without adequate follow-up observations, it is impossible to rule out the BBH scenario at this stage.
The candidate binary black hole separation is 4.3~pc, as inferred from the angular separation and the galaxy's redshift. Note that the distinction between two separate components in B2\,1059+29 is not very obvious and could also be a result of imaging artifacts.

\subsubsection{J1328+2752}
J1328+2752 is another source that shows dual VLBI components (see \cite{Nandi2021} for a detailed discussion of this source). However, the dual AGN nature has not yet been confirmed due to the lack of spectral index information. \cite{Nandi2017} have pointed out that the source possesses a flat spectrum core based on its VLA image. 
We must obtain near-simultaneous observations at dual frequencies to confirm the nature of both the VLBA components. The projected separation between the two components is equal to 6~pc \citep{Nandi2021}.

\subsubsection{4C\,+61.23}
\cite{Liu2018} have presented the 8.4 GHz VLBA image of this AGN, which shows a double structure. Our follow-up observations at 15~GHz are similar to the 8.4~GHz image but suffer from low SNR. One of the dual VLBI components appears extended and might be a core-jet structure by itself, while the south-eastern component might be another AGN core. It is also possible that either component is an AGN core, and the rest of the emission is jet-related. Our follow-up uGMRT observations show a dimmed core at the 400~MHz image compared to L-band ($\alpha\sim-0.25$). The angular separation between the dual VLBA components would correspond to a projected distance of 4.6~pc at the source's redshift. 
 
\subsection{Radio Spectral Index Analysis}

As our sample galaxies have multi-band radio data, we have calculated the spectral indices and used those to obtain the minimum magnetic field and approximate ages for both the primary and secondary lobes.
First, we mask the nuclear emission using the circular aperture from {\it photutils}\footnote{https://photutils.readthedocs.io/en/stable/aperture.html} then we estimated the flux densities from primary and secondary jet (by interactively defining regions using {\tt CASA}). The errors in flux densities are calculated using $\sqrt{(\sigma_\mathrm{rms}\cdot\sqrt{N_b})^2 + (\sigma_p\cdot S)^2}$ \citep{Kale2019} where $\sigma_\mathrm{rms}$ is the rms noise of the maps, N$_\mathrm{b}$ is the number of beams inside the selected regions, S is the sources flux density and $\sigma_\mathrm{p}$ is the absolute flux density scale error in percentage which is assumed to be 0.1 (conservative for the GMRT, VLA and VLBA).
We assume that the total energy of the synchrotron emitting plasma in the jet is close to the equipartition of the magnetic field and the energy in particles \citep{Burbidge1959}. Using the following expressions from \citet{Odea1987} we calculate the total luminosity (L$_{rad}$) and minimum magnetic field (B$_{min}$): 
\begin{equation}
L_{\text{rad}} = 1.2 \times 10^{27} D_L^2 S_0\frac{ (\nu_u^{1 + \alpha} - \nu_l^{1 + \alpha})}{\nu_0^{\alpha} (1 + \alpha)(1 + z)^{1 + \alpha}}
\end{equation}
\begin{equation}
B_{\text{min}} = \left[ \frac{2\pi (1 + k) c_{12} L_{\text{rad}}}{(V\phi)} \right]^{2/7}
\end{equation}

Here, D$_L$ is the luminosity distance in Mpc, and S$_0$ is the total flux in Jy at the frequency $\nu_{0}$ in Hz. We have taken the radio spectrum range from 10 MHz ($\nu_l$) to 100 GHz ($\nu_u$). Assuming that the relativistic electron and proton have a similar energy, we have set the ratio $k$ = 1, and the constant c$_{12}$ is obtained from \citet{Pacholczyk1970} depending on the calculated spectral index values and the frequency cutoffs. The jets are assumed to have cylindrical symmetry, and hence their volumes are calculated as V=$\pi(w/2)^{2}l$ where $w$ is the width in kpc, and $l$ is the length of the jet in kpc. The jets are filled with relativistic particles and magnetic fields uniformly, i.e., $\phi$=1.

Next, we estimated the lifetime of electrons in the radio jet using the following relation from \citet{vanderlaan1969} where the electrons are losing energy via synchrotron as well as inverse- Compton emission mechanisms due to CMB (cosmic microwave background) photons:

\begin{equation}
t =   \frac{2.6 \times 10^4 {B_{min}}^{1/2}}{{({B_{min}^2 + {B_R})^2}[(1+z)\nu_b]^{1/2}}} 
\end{equation}

, where B$_{min}$ is the minimum magnetic field calculated above, assuming the equipartition assumption, $\nu_b$ is the break frequency, and B$_R$ is the magnetic field equivalent to the radiation via IC-CMB. The estimated electron age and other parameters in this section are given in Table~\ref{tab:spectral aging}. 

The primary lobe is steeper than the secondary lobe in 4C\,61.23. However, the difference is smaller than the quoted error bars. The spectral ages of the primary lobes appear to be larger than the secondary lobes in both 4C\,61.23, and 4C\,32.25. According to the most well-accepted models, including backflow diversion and precessional models, the primary lobes are expected to comprise the younger relativistic plasma population. This discrepancy could be because of the assumptions involved in calculating spectral age. Since we only have data at two frequencies, we assumed the highest frequency of our observations to be the break frequency. Moreover, the relativistic plasma need not follow the equipartition assumption.
Hence, we abstain from making further comments in the absence of data at higher frequencies.

\begin{center}
\begin{table*}
\caption{The estimations from the equipartition assumption: $L_{\text{rad}}$, $B_{\text{min}}$, Age}
\begin{tabular}{cccccccc}
\hline \hline
Source & Reference & Flux$_{\text{Ref freq}}$ & $\alpha$ & L$_{\text{rad}}$ & B$_{\text{min}}$ & Age \\
& Frequency (GHz)& (Jy)&  &(erg s$^{-1}$) &  ($\mu$G)&  (Myr)\\
\hline

4C\,32.25 (primary)  & 1.4  & 1.69$\pm$0.16 & -0.43$\pm$0.05 & (4.91$\pm$0.97)$\times 10^{42}$ & 2.17$\pm$0.12 & 51.10$\pm$1.44\\
4C\,32.25 (secondary)  & 1.4 & 0.26$\pm$0.03 & -0.55$\pm$0.08  & (1.09$\pm$0.26)$\times 10^{41}$ & 0.40$\pm$0.02 & 22.15$\pm$0.77\\
4C\,61.23  (primary) & 1.25 & 0.91$\pm$0.09 & -0.58$\pm$0.08  & (1.14$\pm$0.24)$\times 10^{42}$ & 1.72$\pm$0.10 & 37.52$\pm$1.14\\
4C\,61.23 (secondary) & 1.25 & 0.38$\pm$0.03 & -0.52$\pm$0.06  &(1.36$\pm$0.23)$\times 10^{42}$ & 1.16$\pm$0.05 & 30.89$\pm$0.74\\
B2\,1059+29     & 1.25  & 0.13$\pm$0.13 & -0.42$\pm$0.05  & (2.29$\pm$0.37)$\times 10^{42}$ & 2.93$\pm$0.13 & 50.01$\pm$1.16\\

\hline
\label{tab:spectral aging}
\end{tabular}
\end{table*}
\end{center}

\subsection{BH Mass Estimates}
\label{sec:bhestimates}
We estimate the total black hole mass using the $M-\sigma_\star$ relation \citep{Gebhardt2000, Ferrarese2000, Tremaine2002}.

\begin{equation}
M_{\rm BH}=10^{8.13\pm0.06}\left(\frac{\sigma}{200\, \rm km\,s^{-1}}\right)^{4.02\pm0.32}M_{\odot}
\label{eq:3}
\end{equation} 

The velocity dispersion of the target galaxies is estimated by fitting the absorption line spectra via direct pixel method \citep{Ge2012}. We have tabulated the stellar velocity dispersion values and the BH masses for the five sources in our sample in Table~\ref{tab:BH-mass-estimates-nelson}. The ratio of the masses of individual black holes can be approximately determined following Kepler's law by the below equation.

\begin{equation}
   \frac{M_{\rm BH1}}{M_{\rm BH2}} \sim \frac{v_{2}}{v_{1}}
\end{equation}
Here $v_{1}$ and $v_{2}$ are the line of sight rotation velocity of each peak. 
The BH mass ratios are $M_{\rm BH1}/M_{\rm BH2}$ $\sim$ 1.97, 0.67 and 1.36 for galaxies 4C+32.25, B2\,1059+29 and 4C+61.23 respectively.


\cite{Nelson2000} studied the relationship between the [O~{\small III}] $\lambda$5007 line widths and the black hole masses for AGN, under the assumption that the forbidden line kinematics is determined by the gravitational influence of the galaxy bulge. 
They find a moderately strong correlation between the [O {\small III}] line widths and black hole masses.


\begin{equation}
\rm log_{10}(M_{\rm BH})=(3.7\pm0.7)\, \rm log_{10}\left( \frac{FWHM [O{\small\,III}]}{2.35} \right)-(0.5\pm0.1)
\label{eq:2}
\end{equation}  
We use the above equation to determine the individual black hole masses, assuming that separate NLR regions produce these lines. The FWHM of the blueshifted and redshifted [O {\small III}] narrow emission lines were obtained from \cite{Ge2012}. The individual black hole masses obtained using equation~\ref{eq:2} and the total mass estimated from both the relations match within error bars (see Table~\ref{tab:BH-mass-estimates-nelson}). The precessional period estimated from our radio data can now be applied to obtain the separation between the two black holes, assuming that a dual AGN at the center is responsible for both the double-peaked emission lines and the X-shaped radio morphology.

\begin{table*}
\tiny
\caption{Black hole mass estimates from emission lines }
\tabcolsep=0.11cm
\begin{tabular}{cccccccccccc}
\hline \hline
Source & $\Delta~v_b$& $\Delta~v_r$& $\sigma_{*}$ & $\sigma_{\rm b,[O\,III]}$ & $\sigma_{\rm r,[O\,III]}$ & log($M_{\rm BH2}$) & log($M_{\rm BH1}$) & log($M_{\rm tot,\sigma_{*}}$) & log($M_{\rm tot,sum}$) & $M_{\rm BH}$ ratio & $M_{\rm BH}$ ratio \\
name &(km s$^{-1}$) &(km s$^{-1}$) & (km s$^{-1}$) & (km s$^{-1}$) & (km s$^{-1}$) & (M$_{\odot}$) & (M$_{\odot}$) & (M$_{\odot}$) & (M$_{\odot}$) & $\frac{vr}{vb}$ & $\frac{M_{\rm BH2}}{M_{\rm BH1}}$ \\
(1) & (2) & (3) & (4) & (5) & (6) & (7) & (8) & (9) & (10)  & (11)& (12)\\
\hline

4C +32.25 & -211.0$\pm$9.30&107.2$\pm$12.80 & 222.7 $\pm$ 5.3 & 105.2$\pm$7.6 & 165.6$\pm$10.5 & 7.0$\pm$1.4 & 7.7$\pm$1.6 & 8.32$\pm$0.07 & 7.8$\pm$1. & 0.50 $\pm$0.06& 0.19$\pm$0.09 \\
3C 223.1 & -148.5$\pm$5.00&131.2$\pm$3.90 & 200.2$\pm$7 & 112$\pm$2.9 & 121.3$\pm$2.4 & 7.1$\pm$1.4  & 7.2$\pm$1.5 & 8.13$\pm$0.09 & 7.5$\pm$1.5 & 0.88$\pm$0.04 & 0.79$\pm$0.1\\
B2 1059+29 &-167.9$\pm$5.50&249.3$\pm$5.80 &  215.2 $\pm$ 8.0 & 164.2$\pm$5.3 & 124.5$\pm$5.1  & 7.7$\pm$1.6 & 7.3$\pm$1.5 & 8.26$\pm$0.09 & 7.8$\pm$1.5 & 1.48$\pm$0.06& 2.51$\pm$0.8\\
4C +61.23 & -39.2$\pm$1.90&261.4$\pm$2.10 & 200.3 $\pm$ 11.4 & 139.6$\pm$1.1 & 102.6$\pm$1.2  & 7.4$\pm$1.5 & 6.9$\pm$1.4 & 8.13$\pm$0.12 & 7.6$\pm$1.5 & 6.66$\pm$0.32 & 3.16$\pm$0.7 \\
J1430+5217$^a$ & -191.7$\pm$5.46&169.$\pm$7.14 & 251.6$\pm$22.0 & 132.5$\pm$4.8 & 105.2$\pm$6.25 & 7.4$\pm$1.5 & 7.0$\pm$1.4 & 8.53$\pm$0.17 & 7.5$\pm$1.5 & 0.88$\pm$0.04 & 2.5$\pm$0.7\\
\hline
\label{tab:BH-mass-estimates-nelson}
\end{tabular}
\parbox{\textwidth}{\textbf{Note:} Column 1: Source names. J1328+2758 is not presented in this paper, and we refer the reader to \cite{Nandi2021} for details on this source. Columns 2 and 3: Velocity offset of the blue and the blue component from the systemic velocity. Column 4: Stellar velocity dispersion. Column 5: Blue narrow line component velocity dispersion. Column  6: Red narrow line component velocity dispersion. Columns 7 and 8: Logarithm of black hole mass for the blue and red components, respectively (see Section~\ref{sec:bhestimates} for more details). Column 9: Logarithm of the total black hole mass from M-$\sigma_{*}$ relation. Column 10: Logarithm of the total black hole mass estimated as a sum of the individual black holes. $^a$-The emission line properties for all sources except J1430+5217 were obtained from \cite{Ge2012} and \cite{Wang2009}. We have completed the double Gaussian model fitting for J1430+5217 and presented the details in Section~\ref{sec:specmodel}. Column 11: Black hole mass ratio estimated using the ratio $v_r/v_b$. Here, $v_r$ and $v_b$ represent the offset of the emission lines' peak from the galaxy's systemic velocity. Column 12: Black hole mass ratio estimated directly using columns 7 and 8. Since the black hole masses depend on the FWHM of the individual line profiles, this estimation method is independent of the method used for Column 11.}
\end{table*}

\subsection{Precession Modeling}\label{sec:PrecModFit}


We applied a precession model to the jet/counter-jet of the radio sources of our sample. In summary, this model assumes that the jet propagates with a constant speed $v=\beta c$ ($0<\beta<1$), making an instantaneous angle $\phi$ with the line of sight. Due to precession, jet/counter-jet will form an apparent cone-like structure with a semi-aperture angle $\varphi_0$. The precession cone axis makes a viewing angle $\phi_0$, and it is projected on the sky plane with a position angle $\eta_0$ (measured positively from North to East direction). The initial jet direction is reached after an elapsed time $P_\mathrm{prec,obs}$, the precession period measured at the observer's reference frame. The two additional free parameters are the sense of precession, $\iota$ (=1 for clockwise sense and $\iota=-1$ for counter-clockwise sense), and the precession phase $\tau_\mathrm{0,s}$. For further details on the kinematic precession model adopted in this work, see, e.g., \citet{Caproni2009}.

We proceeded with a systematic search for the most suitable precession model parameters for each source of our sample, except for J1328+2758, for which the jet precession analysis was previously done in \citet{Nandi2021}. These parameters are listed in \autoref{tab:PrecParam}, while the precession helices generated from these models are shown in \autoref{fig:precmodbest}. 
The precession helices are produced under the assumption that the jet and counter-jet are formed by plasma elements ejected steadily from the core during a certain interval and with a constant speed (e.g., \citealt{Nandi2021}). A given plasma element ejected at a time $t_\mathrm{ej}$ and with a proper motion $\mu(t_\mathrm{ej})$ will present right ascension and declination offsets from the core at a time $t_\mathrm{obs} (\ge t_\mathrm{ej})$, $\Delta RA(t_\mathrm{obs})$ and $\Delta DEC(t_\mathrm{obs})$ respectively, calculated from (e.g., \citealt{Caproni_Abraham_2004, Caproni2009}):

\begin{equation} \label{DRA_prec}
\Delta RA(t_\mathrm{obs}) = \pm\mu(t_\mathrm{ej})\sin\left[\eta(t_\mathrm{ej})\right]\Delta t_\mathrm{obs},
\end{equation}

\begin{equation} \label{DDec_prec}
\Delta DEC(t_\mathrm{obs}) = \pm\mu(t_\mathrm{ej})\cos\left[\eta(t_\mathrm{ej})\right]\Delta t_\mathrm{obs},
\end{equation}
\\where $\Delta t_\mathrm{obs} = t_\mathrm{obs}-t_\mathrm{ej}$ and the signs `+' and `-' refer respectively to the jet and counter-jet. 

The variables $\mu$ and $\eta$ are calculated from equations (8)--(15) in \citet{Caproni2009}, while $\Delta t_\mathrm{obs}$ is derived from equations (16)\footnote{The sign `-' in this equation (Doppler boosting factor) must be replaced by the sign `+' for the counter-jet.} and (17) in \citet{Caproni2009} as well. More details concerning the jet precession model for each radio galaxy are provided below, including the acceptable ranges for each individual jet precession parameter (estimated from visual inspection after varying one of the free model parameters, keeping the remaining ones fixed).

\subsubsection{J1430+5217}\label{sec:PrecParam J1430+5217}

\begin{figure*}
    \centering
    \begin{subfigure}{0.4\textwidth}
       \centering         \includegraphics[width=\columnwidth]{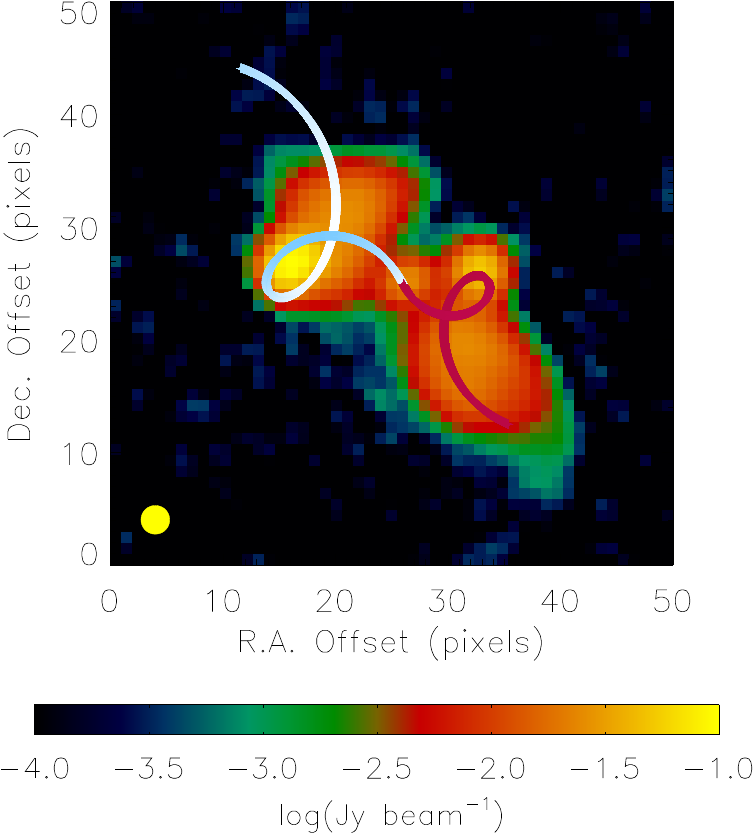}
    \end{subfigure}
     \hfill
     \begin{subfigure}{0.4\textwidth}
       \centering         \includegraphics[width=\columnwidth]{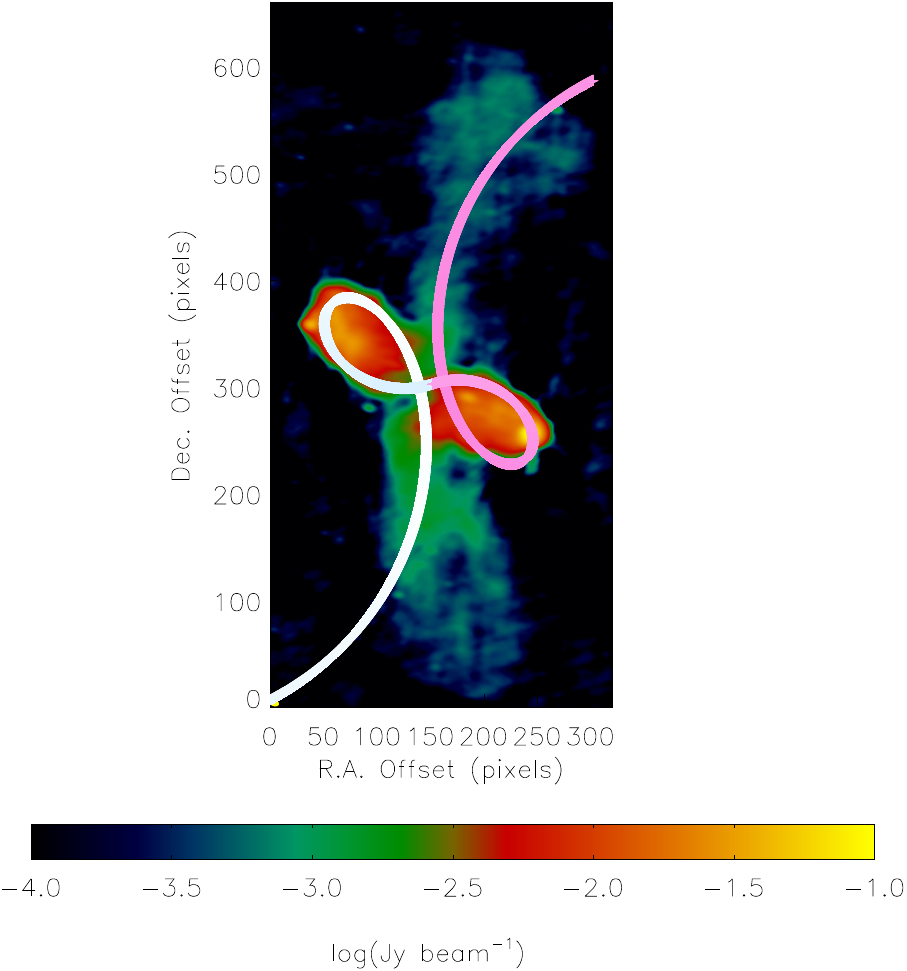}
     \end{subfigure}
     \hfill
     \begin{subfigure}{0.4\textwidth}
       \centering         \includegraphics[width=\columnwidth]{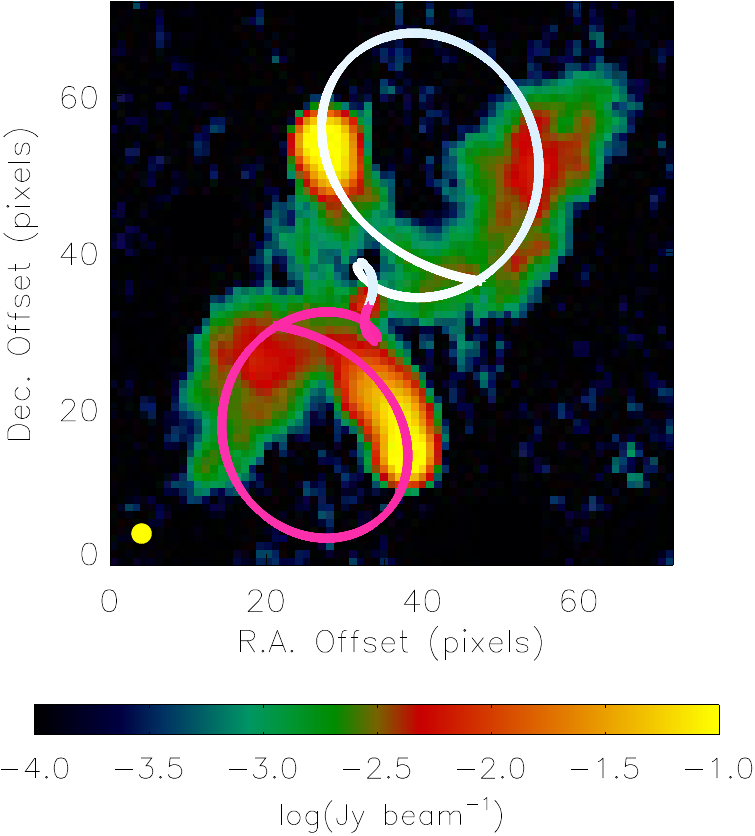}
     \end{subfigure}
     \hfill
     \begin{subfigure}{0.4\textwidth}
       \centering         \includegraphics[width=\columnwidth]{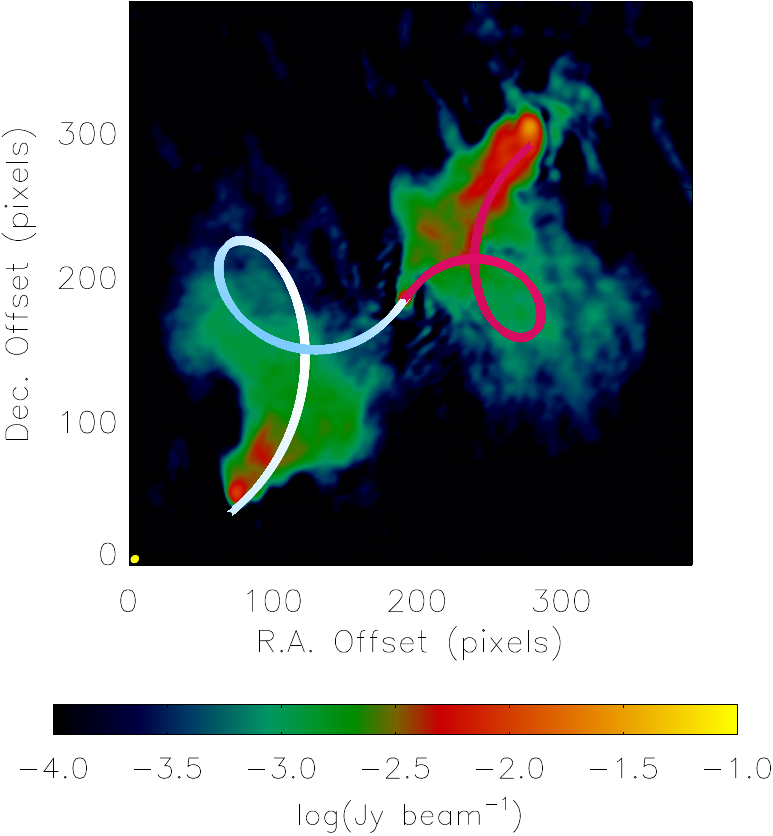}
     \end{subfigure}
     \hfill
     \caption{Superposition of the jet and counter-jet precession helices (blue and purple hues, respectively) generated from the parameters listed in \autoref{tab:PrecParam} on the kiloparsec-scale images of J1430+5217 (1.4 GHz, 1 pixel = $1\farcs8$, upper-left panel), 4C\,+32.25 (0.4 GHz, 1 pixel = $1\farcs3104$, upper-right panel), 3C\,223.1 (1.4 GHz, 1 pixel = $1\farcs8$, lower-left panel), and 4C\,+61.23 (1.25 GHz, 1 pixel = $0\farcs5796$, lower-right panel)
     Lighter colors used in the precession helices represent higher Doppler boosting factors. The yellow ellipse in the lower-left corner of the individual panels represents the FWHM of the elliptical synthesized CLEAN beam.}
    \label{fig:precmodbest}
\end{figure*}

The upper-left panel of \autoref{fig:precmodbest} shows the brightness distribution at 1.4 GHz of the quasar J1430+5217. The brighter regions were labeled as east (E) and west (W) lobes, while more diffuse and extended regions were labeled as north (N) and south (S) wings \citep{Lal2019}. Moreover, the northeast (NE) jet of J1430+5217 is shorter in length than the southwest (SW) jet, even though the former is brighter, in contradiction with the usual behavior of the jet being more extended and brighter than the counter-jet. Precession helices generated from the set of parameters listed in \autoref{tab:PrecParam} are superposed on the 1.4~GHz map of J1430+5217, with the jet pointing towards the NE-direction. This precession model recovers quite well the position of the E and W lobes, as well as the nowadays position angle of its parsec-scale jet seen in \autoref{vlba} ($\sim30\degr$, measured from N to E direction). Doppler boosting factor for the jet (counter-jet) ranges from 1.198 to 1.282 (0.778 to 0.813). Even though our non-accelerating precession model overestimates the total length of the jet at 1.4~GHz, any deceleration of the jet material due to its propagation in a denser environment could bring the tip of the jet helix closer to the N wing.

Acceptable precession models are found for $4.0\leq P_\mathrm{prec,obs}(\mathrm{Myr})\leq 10.0$. Periods shorter than 4.0 Myr put the precession loop seen in \autoref{fig:precmodbest} between the core and W lobe, besides decreasing substantially the precession amplitude. On the other hand, precession periods longer than 10 Myr lead to a loop's position farther than the W lobe. For the remaining model parameters, we obtained the ranges $0.125\leq\beta\leq 0.312$, $20\degr\leq\phi_0\leq32\degr$, $8\degr\leq\varphi_0\leq20\degr$, $54\degr\leq\eta_0\leq85\degr$.

\begin{center}
\begin{table*}
\caption{Model parameters used to generate the precession helices shown in \autoref{fig:precmodbest}.}
\begin{tabular}{cccccccc} \hline \hline
Source & $\iota$ & $P_\mathrm{prec,obs}$ & $\beta$ & $\eta_0$ & $\phi_0$ & $\varphi_0$ & $\tau_\mathrm{0,s}$\\
 &  & (Myr) &  & (deg) & (deg) & (deg) & \\
\hline
4C\,+32.25 & -1 & 87.5 & 0.033 & 105.0 & 31.0 & 29.0 & 0.0\\
B2\,1059+29 (NE jet)& -1 & 48.0  & 0.120 & 95.0 & 7.0 & 6.9 & 0.69\\
B2\,1059+29 (SW jet)& -1 & 96.0  & 0.016 & 279.0 & 26.0 & 25.9 & 0.69\\
4C\,+61.23 & -1 & 6.5  & 0.193 & 106.0 & 30.0 & 17.0 & 0.95\\
3C\,223.1 & 1 & 8.0  & 0.090 & 327.0 & 20.0 & 15.0 & 0.32\\
J1328+2758 & 1 & 4.8  & 0.170 & 343.0 & 40.0 & 17.0 & 0.88\\
J1430+52 & 1 & 5.0  & 0.250 & 63.0 & 26.0 & 14.0 & 0.54\\
\hline
\label{tab:PrecParam}
\end{tabular}
\end{table*}
\end{center}

\subsubsection{4C\,+32.25}\label{sec:PrecParam4C+32.25}

The complex radio structure observed at kiloparsec-scale radio images of 4C\,+32.25 are essentially formed by brighter east and west radio lobes with hotspots at their ends and fainter and more extended north and west wings (e.g., \citealt{Parma1985}). We modeled these radio structures in terms of a precessing jet with a constant speed of 0.033$c$ and a period of about 87.5 Myr. The remaining precession parameters are given in \autoref{tab:PrecParam}, while the precession helices associated with it are shown in \autoref{fig:precmodbest}. Although this precession model assumes that the east lobe and the south wing belong to the jet, another scenario where the jet would be related to the west lobe and the north wing is also possible. Acceptable precession models can also be found for the ranges $40\leq P_\mathrm{prec,obs}(\mathrm{Myr})\leq 500$, even though $P_\mathrm{prec,obs} > 90$ Myr would imply an age for the wings older than about 70 Myr, age estimated by \citet{Klein1995}, $0.006\leq\beta\leq0.072$, $25\degr\leq\phi_0\leq51\degr$, $23\degr\leq\varphi_0\leq49\degr$, $90\degr\leq\eta_0\leq120\degr$.

Jet precession in 4C\,+32.25 was already invoked in previous works \citep{Parma1985, Klein1995}, even though qualitatively. For instance, \citet{Klein1995} proposed a precession period of 200 Myr, while \citet{Parma1985} suggested the values of 25\degr and 60\degr for the precession angle and the viewing angle of the precession cone axis. These values are compatible with those estimated in this work.

\subsubsection{3C\,223.1}\label{sec:PrecParam3C223p1}
The X-shaped source 3C\,223.1 shows two pairs of structures at kiloparsec-scales: lower brightness surface lobes (labeled as East and West wings in \citealt{Lal2005}) with flatter spectral indices in comparison with the higher brightness surface lobes (labeled as North and South lobes in \citealt{Lal2005}). It implies that North and South lobes are more aged structures than East and West wings \citep{Dennett2002, Lal2005, Lal2007}, imposing an important constraint for the jet precession modeling. Indeed, the precession model for 3C\,223.1 shown in \autoref{fig:precmodbest} respects these findings since the jet (counter-jet) helix crosses the North (South) lobe firstly, reaching the West (East) wing after $\sim3$ Myr for this particular set of parameters. This precession model also predicts the correct orientation of the parsec-scale jet of 3C\,223.1 at 5 GHz (P.A. $\sim -25\degr$; see \autoref{vlba}). Doppler boosting factor for the jet (counter-jet) ranges from 1.075 to 1.094 (from 0.914 to 0.928). Acceptable precession models can also be found for the ranges $1.5\leq P_\mathrm{prec,obs}(\mathrm{Myr})\leq 20.9$, $0.035\leq\beta\leq0.46$, $15\degr\leq\phi_0\leq42\degr$, $10\degr\leq\varphi_0\leq35\degr$, $310\degr\leq\eta_0\leq15\degr$. \citet{Gong2011} also modeled this source in terms of jet precession, determining a precession period of $\sim 1.8$ Myr, $\phi_0\sim36\degr$ and $\varphi_0\sim26\degr$, all of them in the parameters' ranges estimated in this work.

Furthermore, no radio emission is seen between the North lobe and the West wing, as well as between the South lobe and the East wing. It might indicate an intermittent AGN activity in 3C\,233.1 that, in terms of the precession model shown in \autoref{fig:precmodbest}, would represent a substantial decrease of the jet activity during about 3 Myr. Surprisingly, a similar 3-Myr interruption of the jet activity might be associated with the brightness gap seen right above the core (at the location of the small jet precession loop seen in \autoref{fig:precmodbest}), even though the south counterpart of this gap is not observed at 1.4 GHz.

\subsubsection{4C\,+61.23}\label{sec:PrecParam4C+61.23}

This radio galaxy exhibits two hotspots at the end of its pair of kiloparsec-scale lobes, oriented along northwest-southeast (NW-SE) direction on the plane of the sky. Two additional northeast-southwest (NE-SW) oriented structures are continuously connected with the radio lobes that could be formed by backflows from the lobes \citep{Lara2001}. In this work, we investigated the possibility that such complex structures could be mainly produced by precessing jets. 

We show the precession helices in \autoref{fig:precmodbest} generated by the model parameters provided in \autoref{tab:PrecParam}. Overall agreement between model and image can be seen in this figure, suggesting that NE-SW structures could be produced by a precessing jet. Our kiloparsec-scale precession model for 4C\,+61.23 also correctly predicts the current position angle of the parsec-scale jet seen in VLBI images of this source \citep{Liu2018}. Jet is assumed to point towards SE direction, which is corroborated by the more distant location of the SE hot-spot from the core region of 4C\,+61.23 in comparison with the NW hot-spot, as well as from VLBI observations of this source, for which its parsec-scale jet seems to be pointing to southeast \citep{Liu2018}. Note that the SE lobe seems slightly fainter than the NW lobe, contrary to what would be expected in this case, indicating some hydrodynamic effects on the jet propagation that are not considered in our pure kinematic precession model.

Fair precession models can be found for the following parameter ranges: $4.5\leq P_\mathrm{prec,obs}(\mathrm{Myr})\leq 70.0$, $0.018\leq\beta\leq 0.279$, $25\degr\leq\phi_0\leq35\degr$, $12\degr\leq\varphi_0\leq22\degr$, $100\degr\leq\eta_0\leq111\degr$.

\begin{figure*}
    \centering
    \begin{subfigure}{0.45\textwidth}
       \centering         \includegraphics[width=\columnwidth]{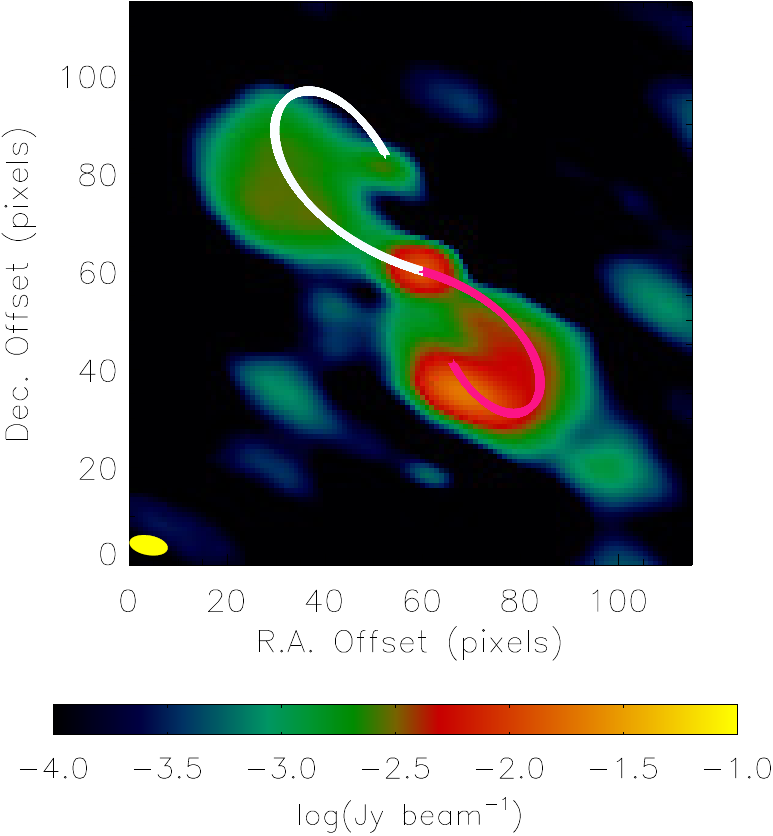}
    \end{subfigure}
     \hfill
     \begin{subfigure}{0.45\textwidth}
       \centering         \includegraphics[width=\columnwidth]{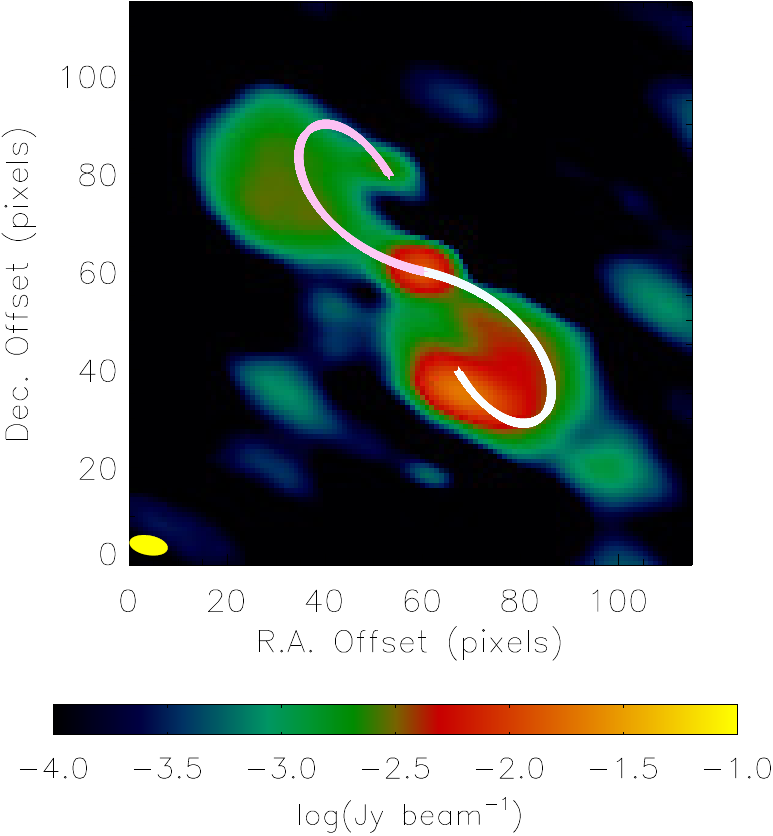}
     \end{subfigure}
     \hfill
     \caption{Superposition of the jet and counter-jet precession helices (blue and purple hues, respectively) generated from the parameters listed in \autoref{tab:PrecParam} on the kiloparsec-scale image of B2\,1059+29 (1.25 GHz, 1 pixel = $0\farcs5508$). Given the contradictory findings regarding surface brightness and total length of the pair of lobes, we have considered two distinct scenarios where that jet points towards the NE direction (left panel) and the SW direction (right panel). Lighter colors used in the precession helices represent higher Doppler boosting factors. The yellow ellipse in the lower-left corner of the individual panels represents the FWHM of the elliptical synthesized CLEAN beam.}
    \label{fig:precmodbestB21059+29}
\end{figure*}

\subsubsection{B2\,1059+29}\label{sec:PrecParamB21059+29}

This radio galaxy shows an S-symmetry morphology at the Band-5 map, with the SW lobe brighter than the NE one. It suggests its jet is pointing toward the SW direction. However, the NE structure is more extended than the SW lobe, which contradicts the former possibility. Unfortunately, no parsec-scale image of B2\,1059+29 that could shed some light on this contradictory finding is available in the literature, obligating us to consider both possibilities in this work.

We show in \autoref{fig:precmodbestB21059+29} two viable jet precession models, one associating the NE lobe with the kiloparsec-scale jet of B2\,1059+29 (left panel) and the other with the jet oriented southwestward (right panel). Besides the jet orientation, these two models differ mainly in terms of the precession period, having the SW model a precession period twice longer than that assumed in the NE precession model. Even though the NE jet precession model seems to be slightly better in comparison with the SW model due to the NE precession helix fully covers the total extension of the NE lobe of B2\,1059+29, a roughly W-oriented motion of the galaxy might boost the brightness of the SW lobe, as well as increase (decrease) the length of the NE (SW) lobe. In this case, the SW model could be preferable. This possibility will be analyzed in more detail using hydrodynamic simulations in future work.

Precession model parameters that produce fair agreements with the radio structures of B2\,1059+29 are $25.0\leq P_\mathrm{prec,obs}(\mathrm{Myr})\leq 140.0$, $0.041\leq\beta\leq 0.230$, $5.8\degr\leq\phi_0\leq21\degr$, $5.7\degr\leq\varphi_0\leq20.9\degr$, $85\degr\leq\eta_0\leq105\degr$ in the case of NE-oriented jet. In contrast, the ranges $40.0\leq P_\mathrm{prec,obs}(\mathrm{Myr})\leq 180.0$, $0.008\leq\beta\leq 0.037$, $10.0\degr\leq\phi_0\leq41\degr$, $9.9\degr\leq\varphi_0\leq40.9\degr$, $269\degr\leq\eta_0\leq285\degr$ are applicable for the case of a jet oriented southwestward.

\section{Discussion}
\label{sec:disc}

In this section, we explore the three different signatures that point to the presence of BBH and the feasibility of a BBH model. We will discuss each signature independently and then consider how they compare with each other for a comprehensive understanding.

\subsection{Role of BBH in double-peaked emission lines }
\label{subsec:dpagnbbh}
Assuming that the double-peaked [O\,III] emission line has its origin from the binary black holes, it is possible to calculate the separation between the black holes, $d_\mathrm{BH}$, through

\begin{equation} \label{dBH}
d_\mathrm{BH} = \frac{GM_\mathrm{tot}}{\Delta v^2},
\end{equation}
\\where $G$ is the gravitational constant and $\Delta v$ is the velocity separation of the blueshift and redshift line components.

Using the values of $M_\mathrm{tot}$ given in \autoref{tab:BH-mass-estimates-nelson}, as well as the measured velocity separation between the [O III] emission lines, we obtain for $d_\mathrm{BH}=11.8\pm4.6$ pc for J1430+5217, $8.9\pm1.7$ pc for 4C\,+32.25, $7.1\pm1.5$ pc for 3C\,223.1, $6.4\pm1.8$ pc for 4C\,+61.23, and $4.5\pm0.9$ pc for B2\,1059+29.

The size of the NLR can be obtained using the relation derived in \cite{Liu2013},

   

\begin{multline}
    \rm log \left(\frac{R_{\rm NLRs}}{pc}\right) = (0.250 \pm 0.018)
    \times \, log \left( \frac{L_{\rm [O III]}}{10^{42} \, \rm erg \, s^{-1}} \right) \\
    + (3.746 \pm 0.028) \rm  
\end{multline}

We estimate the size of the NLR region for all our sources using the above equation. The [O\,{\small{III}}] luminosities are taken from \cite{Ge2012}.
The sizes of the NLR regions for both the redshifted and blueshifted components were computed separately and are provided in Table~\ref{tab:BHsize}.
\begin{table}
\centering
\caption{NLR region sizes for the DP emission lines}
\tabcolsep=0.11cm
\begin{tabular}{ccc}
\hline
\hline
{Source} & {NLR size of blueshifted} & {NLR size of redshifted } \\ 
 name             & {component (kpc)} & {component (kpc)} \\ \hline
4C32.25  & $(1.54\pm0.17)$ & $(1.65\pm0.18)$ \\
3C223.1  & $(3.69\pm0.26)$ & $(4.13\pm0.28)$ \\
B21059+29 & $(2.88\pm0.23)$ & $(2.70\pm0.22)$ \\ 
4C61.23  & $(5.6\pm0.4)$   & $(4.76\pm0.31)$ \\
J1430+5217 & $(5.9\pm0.4)$   & $(5.31\pm0.34)$ \\  \hline
\end{tabular}

\label{tab:BHsize}
\end{table}


To explain the BBH model successfully, the size of the NLR regions ideally should not be greater than BH separations. However, we derive black hole separation much smaller than the estimated NLR sizes in all our sample galaxies. Such a significant disparity in the size scales argues against a scenario where binary black holes could be the primary cause of the double-peaked emission lines. However, it has to be noted that these are merger systems, and the standard empirical relation need not apply in this case. Moreover, while the NLR could extend to sizes as large as a few kpcs, the NLR emission may be more concentrated in the inner pc-scale regions, which are influenced by the individual black holes, possibly leading to the double-peaked emission lines. We cannot spatially resolve the NLR with optical spectroscopy, so the exact nature or the extent of NLR remains uncertain. There may be a complex structure in the NLR \citep{popovic2012}; for instance, the NLR may have been tidally disrupted, leaving only a small core component around each BH \citep{Haas2005,Baum2010}. 
It is interesting to note that the double-peak is present not only in [O\,III] emission lines but also in the H$\beta$ (BLR) regions (see Figure~\ref{fig:J1430+5217} for example). Hence, at this point, it is impossible to rule out the binary black hole scenario completely.

Another likely explanation for the double-peaked emission lines in the spectra is the jet-driven outflows. Given that all our sample galaxies have jets that extend over a few kpcs, it is possible that the gas in the NLR is being pushed away, leading to the double peak seen in the optical spectra \citep{Stockton2007, Kharb2017b, Kharb2019}. The outflow velocities in a jet-driven outflow could also range from 50 to several 100s of km s$^{-1}$ \citep[see for example,][]{Mahony2016}. Hence, it is hard to distinguish between the two possible mechanisms that give rise to the double peak.

Another possible explanation for the origin of the double-peak is peculiar disk kinematics. Owing to the high AGN-related [O\,III] luminosity, which is unlikely to be distributed in a disk. NLR-related [O\,III] emission is often distributed in biconical structures \citep{Falcke1998} rather than a rotating disk. Moreover, the optical hosts of radio-loud sources are early-type galaxies like ellipticals, often lacking a well-defined disk \citep{Best2005}.

Future observations characterizing the spatial distribution of the [O\,{\small{III}}] emission, either using narrow-band imaging or IFS, can help distinguish between the various models. 

\subsection{Precession and the role of BBH}
\label{sec:precession}
The interaction of binary black holes can give rise to inverse symmetric radio jet geometries like X, S, and Z shapes. The peculiar morphology could be either due to jet reorientation or jet precession in the presence of a companion galaxy. There are arguments in the literature suggesting that X or S-shaped morphology can also be due to the back-flow of gas. There are, however, confirmed DAGN (NGC\,326) or binary systems (SS433 microquasar) whose morphologies are well explained by jet precession. \cite{Hjellming1981} proposed a kinetic model to explain the radio jets from X-ray binary SS433 without assuming the origin of jet precession.

\citet{Nandi2021} analyzed three distinct jet precession scenarios in the case of the radio galaxy J1328+2752: misaligned accretion disk with the orbital plane of the binary black holes (e.g., \citealt{Sillanpaa1988, Katz1997, Caproni2017}), geodetic precession (e.g., \citealt{Barker1975, Krause2019}), both scenarios involving supermassive binary black holes, and the Bardeen–Petterson effect (e.g., \citealt{Bardeen1975, Caproni2006, Caproni2006b, Martin2007}), where precession is induced by a single (spinning) black hole. Here, we will follow \citet{Nandi2021}, analyzing the feasibility of these scenarios for the remaining five radio galaxies of our sample that present double-peaked emission lines besides signatures of jet precession. 

\subsubsection{Geodetic precession}
\label{sec:geodeticprecession}
 The black hole separations derived in Section~\ref{subsec:dpagnbbh} imply orbital periods at the observer's reference frame, $P_\mathrm{orb,obs}$\footnote{Calculated using Kepler's third law, i.e., $P_\mathrm{orb,obs} = 2\pi d_\mathrm{BH}^{3/2}/\sqrt{G M_\mathrm{tot}}$.}, equals to $206\pm128$ kyr, $171\pm51$ kyr, $152\pm51$ kyr, $131\pm57$ kyr, and $66\pm22$ kyr, for J1430+5217, 4C\,+32.25, 3C\,223.1, 4C\,+61.23 and B2\,1059+29 respectively.

\begin{figure*}
    \centering
    \begin{subfigure}{0.45\textwidth}
       \centering         \includegraphics[width=\columnwidth]{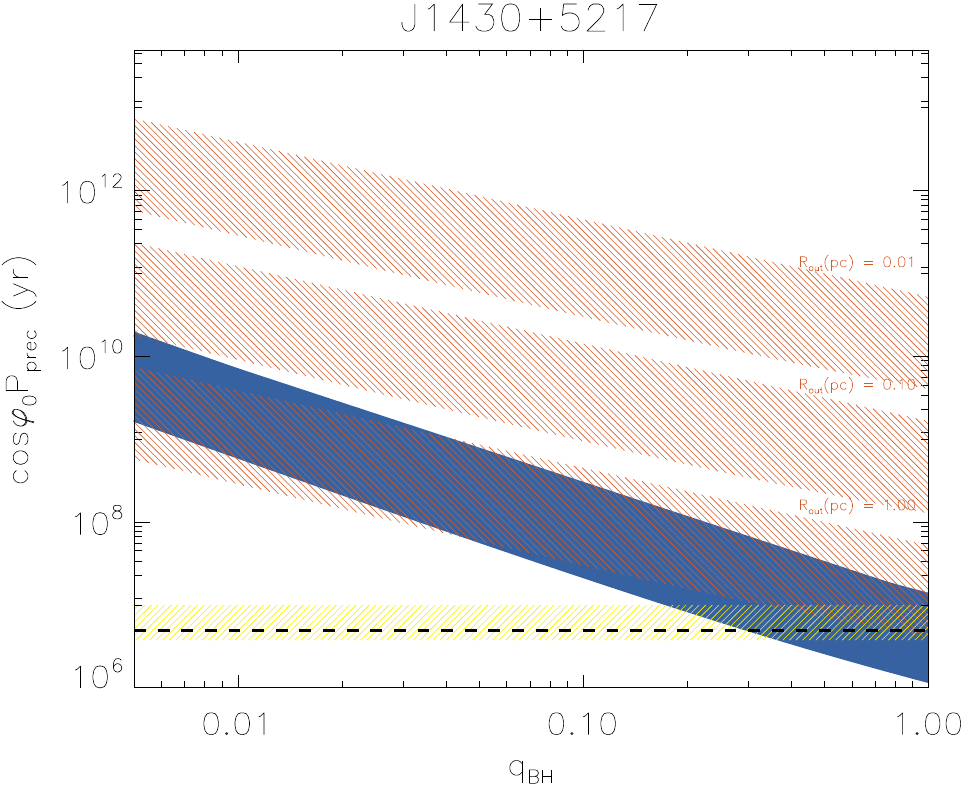}
    \end{subfigure}
     \hfill
    \begin{subfigure}{0.45\textwidth}
       \centering         \includegraphics[width=\columnwidth]{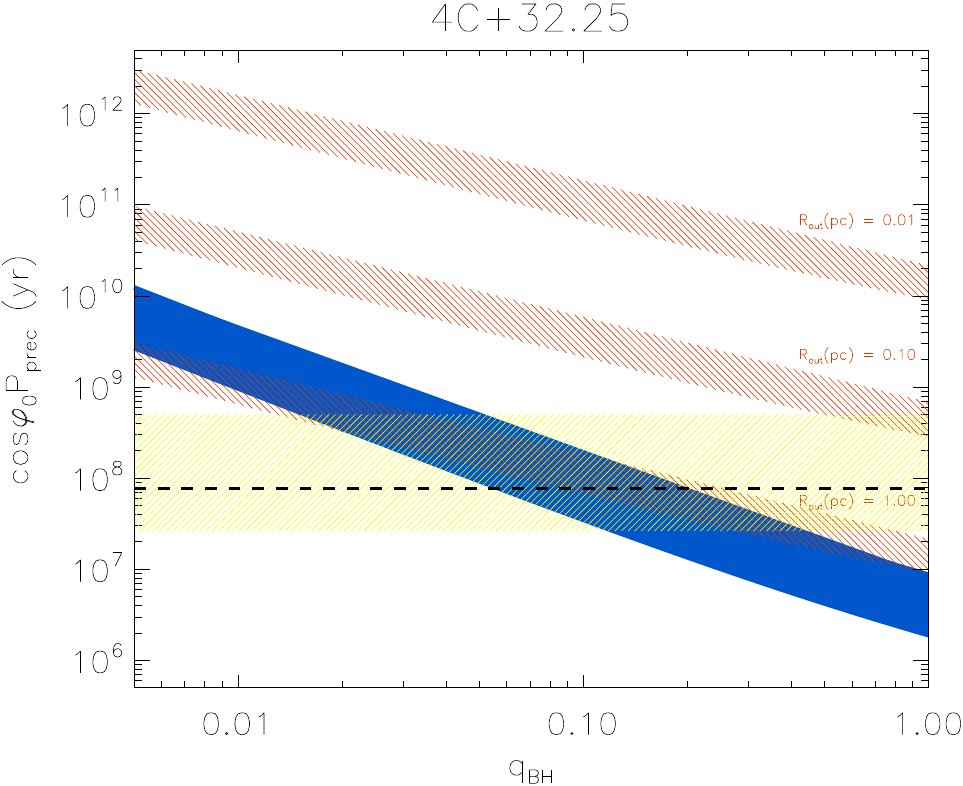}
    \end{subfigure}
     \hfill
     \begin{subfigure}{0.45\textwidth}
       \centering         \includegraphics[width=\columnwidth]{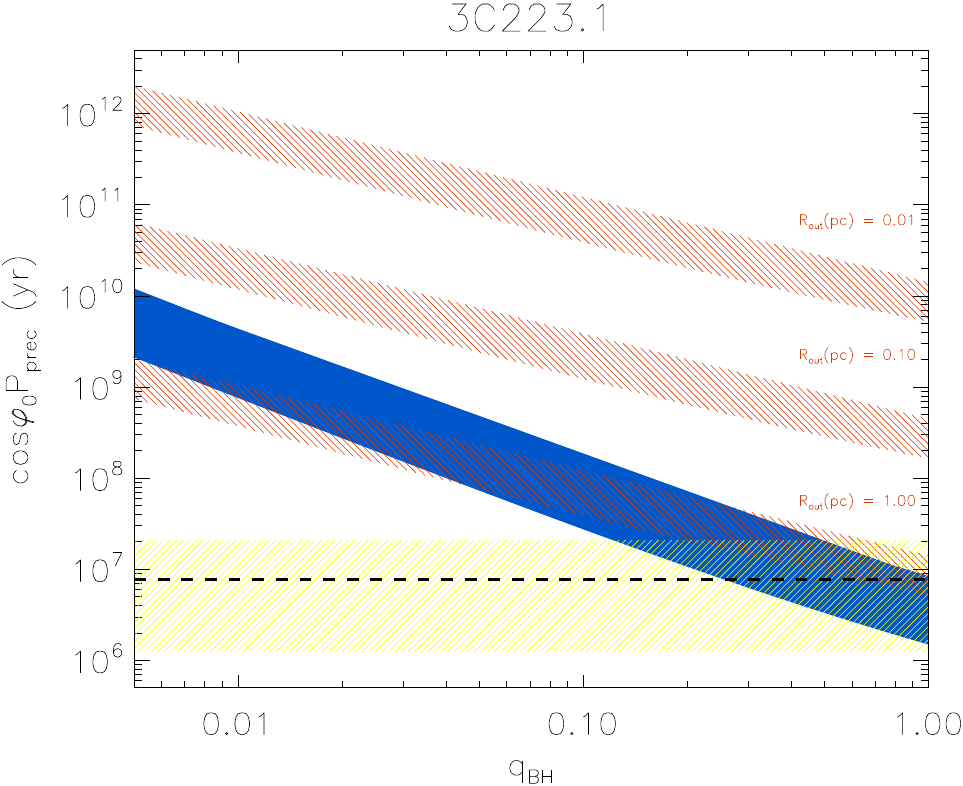}
     \end{subfigure}
     \hfill
     \begin{subfigure}{0.45\textwidth}
       \centering         \includegraphics[width=\columnwidth]{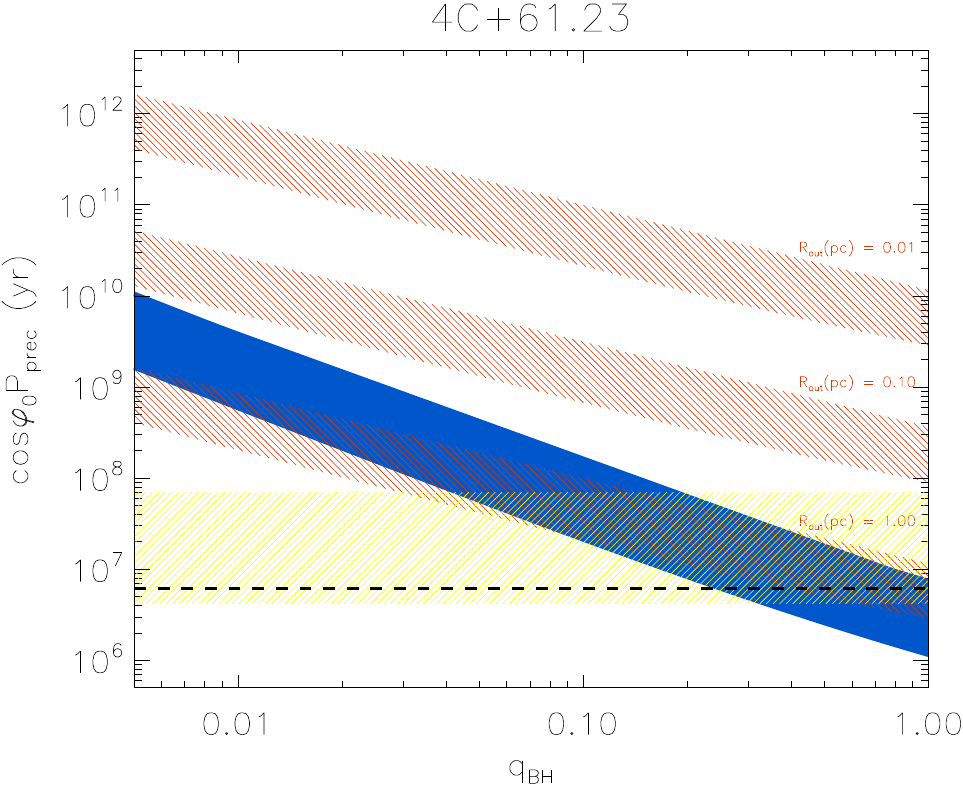}
     \end{subfigure}
     \hfill
     \begin{subfigure}{0.45\textwidth}
       \centering         \includegraphics[width=\columnwidth]{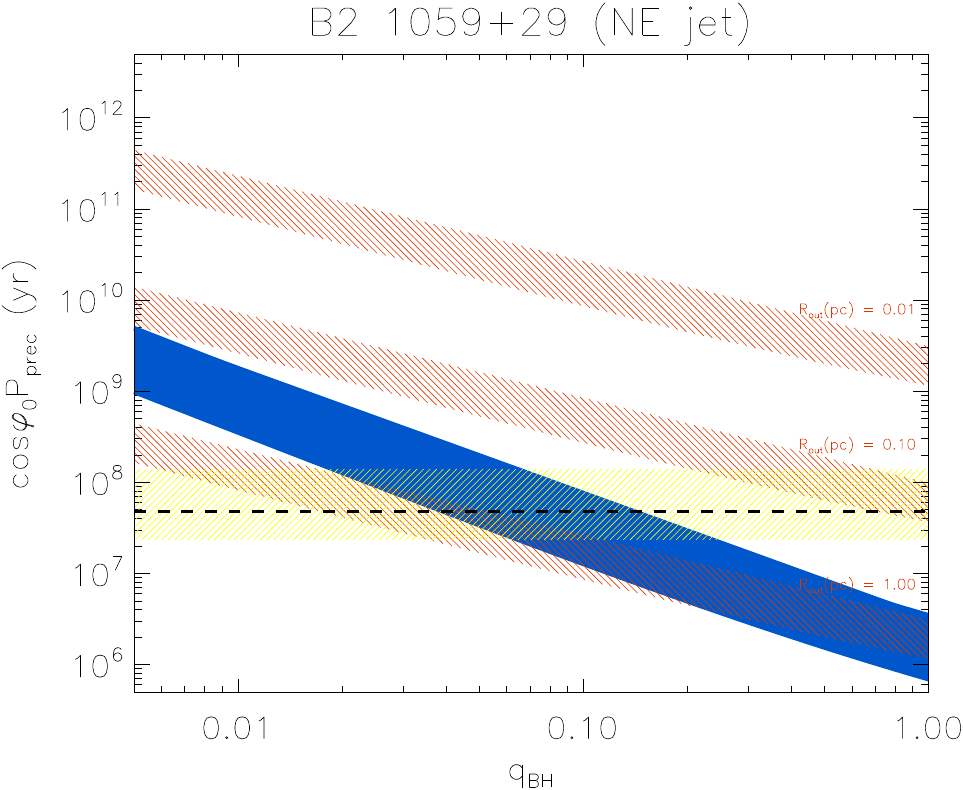}
     \end{subfigure}
     \hfill
     \begin{subfigure}{0.45\textwidth}
       \centering         \includegraphics[width=\columnwidth]{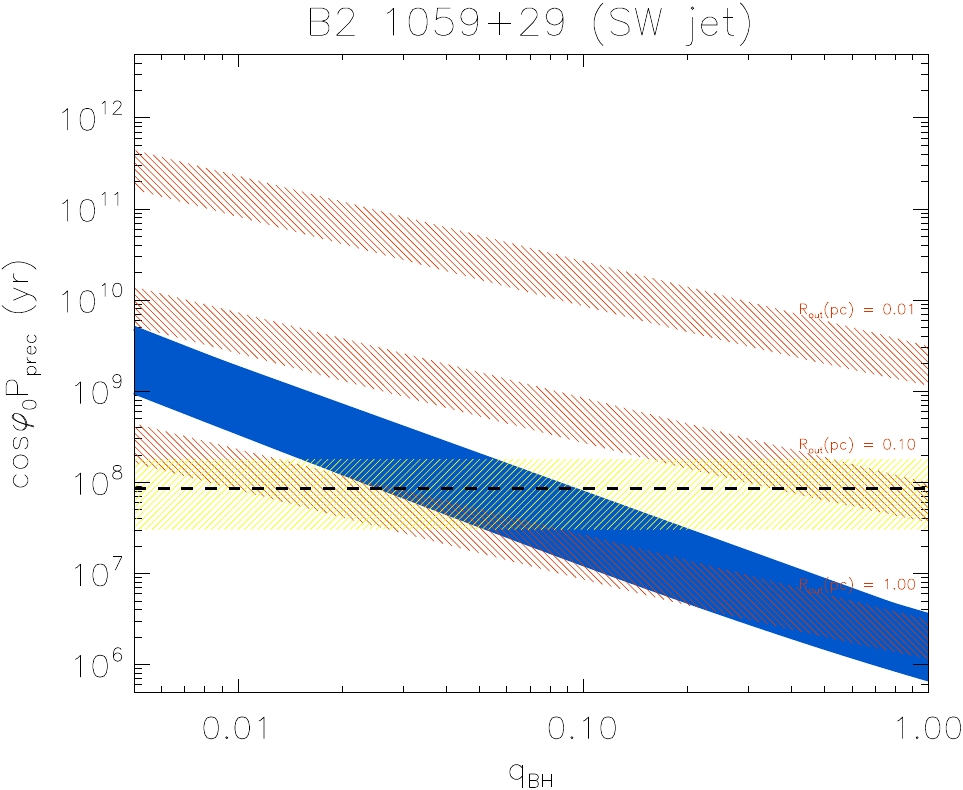}
     \end{subfigure}
     \hfill
\caption{Behaviour of $\cos\varphi_0 P_\mathrm{prec,obs}$ as a function of $q_\mathrm{BH}$ due to precession of the primary’s accretion disc induced by a secondary black hole in a non-coplanar orbit around the primary one. Blue stripes refer to $\cos\varphi_0 P_\mathrm{prec,obs}$ at $1\sigma$-level calculated from equation (6) in \citet{Nandi2021} assuming that the outer radius of the disc is the tidal radius. Red stripes use the same equation but with the outer radius of the disc set as the values 0.01, 0.1 and 1 pc. Hatched yellow rectangles represent the lower and upper conservative limits for $\cos\varphi_0 P_\mathrm{prec,obs}$ expected for each source. The dashed horizontal line refers to the jet precession model listed in \autoref{tab:PrecParam}.}
    \label{fig:SMBBHS_Pprecobs_qBH}
\end{figure*}

\begin{figure*}
    \centering
    \begin{subfigure}{1.0\textwidth}
       \centering         \includegraphics[width=\columnwidth]{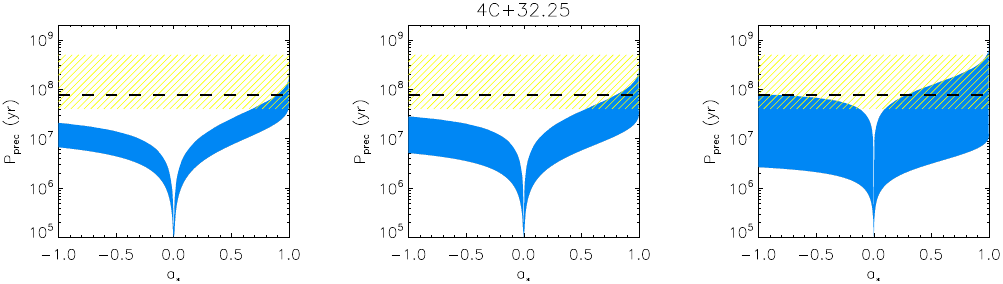}
    \end{subfigure}
     \hfill
    \begin{subfigure}{1.0\textwidth}
       \centering         \includegraphics[width=\columnwidth]{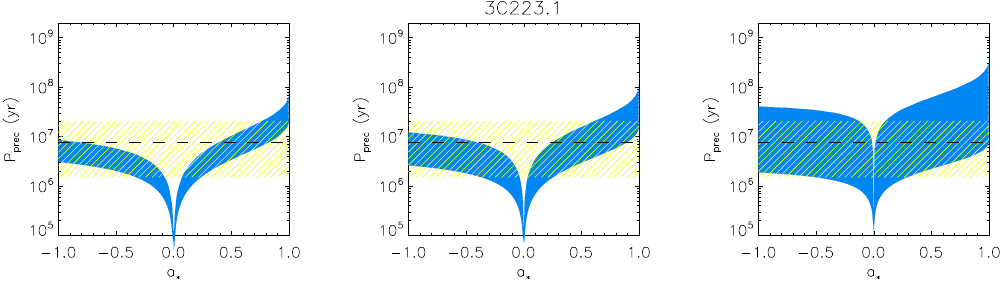}
    \end{subfigure}
     \hfill
     \begin{subfigure}{1.0\textwidth}
       \centering         \includegraphics[width=\columnwidth]{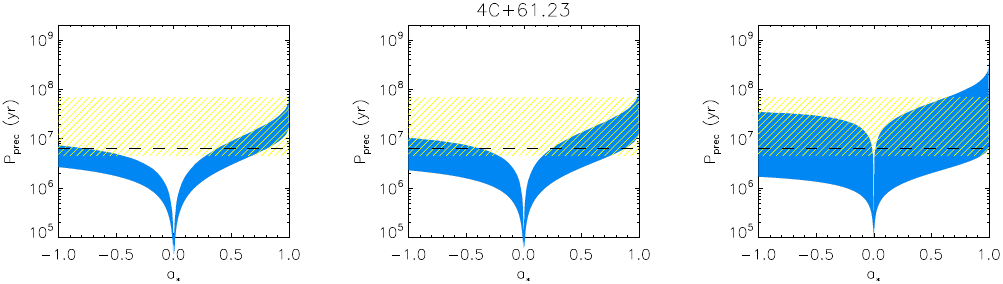}
     \end{subfigure}
     \hfill
\caption{Precession period induced by the BP effect as a function of $a_\ast$ for 4C\,+32.25, 3C\,223.1 and 4C\,+61.23 (top to bottom panels, respectively). It was assumed in the calculations $\alpha = 0.1$ and $R_\mathrm{out,p}=0.05$ parsec. Solutions for $s=-2$, -1, and 0 are shown from left to right. Upper and lower envelopes of the blue stripes for $P_\mathrm{prec,BP}$ refer respectively to the lower and upper values of $\Sigma_0$. The hatched yellow rectangle delineates the lower and upper conservative limits for $P_\mathrm{prec,obs}$ expected for each source.}
    \label{fig:BP_Pprecobs_spin}
\end{figure*}

In the case of the geodetic precession, the minimum value of the precession period, $P_\mathrm{prec}^\mathrm{geo,min}$ is 

\begin{equation} \label{Pprecgeomin}
\left(\frac{P_\mathrm{prec}^\mathrm{geo,min}}{\mathrm{Gyr}}\right)\cong 0.071\left(\frac{d_\mathrm{BH}}{\mathrm{1 pc}}\right)^{5/2}\left(\frac{M_\mathrm{tot}}{\mathrm{10^9 \mathrm{M}_\odot}}\right)^{-3/2}  
\end{equation}
\\which was obtained after assuming $q_\mathrm{BH} = M_\mathrm{BH,s}/M_\mathrm{BH,p} = 1$ in the original equation provided, e.g., by \citet{Krause2019}, where $M_\mathrm{BH,p}$ and $M_\mathrm{BH,s}$ are respectively the mass of the primary (the driver of the jet) and secondary supermassive black holes in a binary system. Using \autoref{Pprecgeomin}, we found $P_\mathrm{prec}^\mathrm{geo,min} = 171\pm166$ Gyr for J1430+5217, $174\pm92$ Gyr for 4C\,+32.25, $191\pm117$ Gyr for 3C\,223.1, $149\pm120$ Gyr for 4C\,+61.23, and  $39\pm24$ Gyr for B2\,1059+29, values incompatible with the precession periods estimated in \autoref{sec:PrecModFit} for these sources. Moreover, these are much higher than the age of the Universe, further invalidating them. It implies that geodetic precession cannot be the mechanism behind the jet precession inferred from our kinematic models.

\subsubsection{Misaligned accretion disk}
Another possibility in the framework of binary systems is the precession induced by a secondary black hole in a non-coplanar orbit around the primary one. If the outer radius of the primary accretion disc is $R_\mathrm{out,p}$, the precession period can be estimated as (e.g., \citealt{Nandi2021})

\begin{equation}\label{PprecSMBBHS}
\left[K(s)\cos\varphi_0\right]\left(\frac{P_\mathrm{prec,obs}}{P_\mathrm{orb,obs}}\right)=\left[\frac{\left(1+q_\mathrm{BH}\right)^{1/2}}{q_\mathrm{BH}}\right]\left(\frac{R_\mathrm{out,p}}{d_\mathrm{BH}}\right)^{-3/2},
\end{equation}
\\where $K(s)\approx 0.19-0.47$ for a power-law surface density accretion disk with an index $s$ between 0 and -2 \citep{Larwood1996, Bate2000}.

We show in \autoref{fig:SMBBHS_Pprecobs_qBH} the precession period at the observer's reference frame times the cosine of the precession angle as a function of  $q_\mathrm{BH}$. Solutions represented by blue stripes were obtained assuming $R_\mathrm{out,p}$ equals the tidal radius of the BBH (e.g., \citealt{Papaloizou1977, Romero2000, Caproni2017}), while those found from a fixed value of $R_\mathrm{out,p}$ (0.01, 0.1 and 1.0 pc, typical values for which AGN accretion discs become gravitationally unstable; e.g., \citealt{Goodman2003, King2007}) are displayed by hatched red stripes. In addition, yellow stripes show the ranges for $\cos\varphi_0 P_\mathrm{prec,obs}$ derived in \autoref{sec:PrecModFit}. Precession induced by a non-coplanar secondary is possible only if $q_\mathrm{BH}\gtrsim 0.012$ for 4C\,+32.25, while $q_\mathrm{BH}\gtrsim 0.029$ is applicable for 4C\,+61.23. For J1430+5217 and 3C\,223.1, the lower limits for $q_\mathrm{BH}$ are respectively 0.18 and 0.13. As the precession modeling for B2\,1059+29 led to two distinct scenarios for its jet precession, we analyzed both possibilities in the context of the present physical mechanism, resulting in the bottom panels in \autoref{fig:SMBBHS_Pprecobs_qBH}. In the case of a jet oriented northeastward, $q_\mathrm{BH}\gtrsim 0.005$ makes this precession mechanism compatible with our jet precession modeling, similar to the results obtained for the jet pointing SW,  $q_\mathrm{BH}\gtrsim 0.004$. Therefore, jet precession induced by torques in the primary accretion disc due to a secondary black hole with an orbit non-coplanar with the primary black hole is a viable mechanism for 4C\,+32.25, 4C\,+61.23, and B2\,1059+29.

\subsubsection{Bardeen Petterson effect}
Finally, disc precession can be driven by the combination of the accretion disc's viscosity and the Lense-Thirring effect \citep{Lense1918} due to the misalignment between the angular momenta of the disc and of a Kerr black hole, a mechanism known in the literature as Bardeen-Petterson effect (BP; \citealt{Bardeen1975}). We analyzed the feasibility of the BP effect to be responsible for precession in the sources 4C\,+32.25, 4C\,+61.23, and 3C\,223.1 only since no information regarding the bolometric luminosity of B2\,1059+29, and J1430+5217 were found in the literature. We follow strictly the formalism adopted in \citet{Nandi2021} (Section 3.3 in their paper), assuming $\alpha=0.1$, the standard viscosity parameter introduced by \citet{Shakura1973}, and an outer radius for the accretion disc of 0.05 parsec. We have made use of the same notation as in \cite{Nandi2021}. The bolometric luminosity, $L_\mathrm{bol}$, for 4C\,+32.25, 3C\,223.1 and 4C\,+61.23 are, respectively, $8.13\times 10^{44}$, $1.15\times 10^{45}$ and $1.34\times 10^{45}$ erg s$^{-1}$ \citep{Trichas2013, Kong2018}, which turns out to be respectively $l=L_\mathrm{bol}/L_\mathrm{Edd} = 0.026, 0.058$ and 0.068 after estimating the Eddington luminosity, $L_\mathrm{Edd}$, from $M_\mathrm{tot}$ given in \autoref{tab:BH-mass-estimates-nelson}.
The results are shown in \autoref{fig:BP_Pprecobs_spin}, considering three different power-law indices, $s$, for the surface density of their accretion discs. 

We can note that the BP effect induces precession timescales compatible with the estimated range for the jet precession period in 4C\,+61.23 for the three values of $s$ ($-2$, $-1$, and 0), even though not for all values of $a_\ast$, a dimensionless parameter corresponding to the ratio between the actual angular momentum of the black hole and its maximum possible value. Indeed, only $a_\ast$ lower than about $-0.6$ (antiparallel angular momentum vectors of the disc and the black hole) or higher than 0.3 (parallel angular momentum vectors of the disc and the black hole) produces precession timescales compatible with those inferred from the radio maps for $s=-2$. At the same time, less restrictive ranges for $a_\ast$ are found for $s=-1$ and 0 (e.g., BP effect is ruled out for $-0.2 < a_\ast < 0.2$ in the case of $s=-1$). Similar behavior was found for 3C\,223.1, but with almost all values of $a_\ast$ and $s$ allowed for this source. For the source 4C\,+32.25, the BP effect is only feasible for prograde rotation with $a_\ast\gtrsim 0.5$ in the case of $s=-2$ and $s=-1$. For a constant surface density profile ($s=0$), retrograde rotation is marginally possible only if $a_\ast \lesssim -0.2$. Therefore, the BP mechanism cannot be ruled out as a potential precession driver in the case of the sources 4C\,+32.25, 3C\,223.1, and 4C\,+61.23.
\subsection{Dual VLBA compact components and the role of BBH}
From our sample of six X-shaped radio galaxies with double-peaked narrow emission lines in optical spectra, we detect double VLBA compact components in three sources. From low-resolution GMRT images, the cores of both 4C\,61.23 and J1328+27 exhibit flat-spectrum, whereas B2\,1059+29 possess a steep spectrum compact component at the centre. A steep core spectral index often indicates that extended structures dominate the emission within the unresolved beam \cite[e.g.,][]{Marecki2023}. Hence, the structure seen in B2\,1059+29 might be a compact symmetric object \cite[CSO;][]{Odea2021}. 

Obtaining quasi-simultaneous multi-frequency observations is crucial to discern the true nature of the components. If both VLBA components are found to possess a flat or an inverted radio spectrum, it would strongly imply that the compact components are AGN cores. One steep spectrum and one flat/inverted spectrum core would indicate a core-jet structure; finally, two steep spectrum components could suggest the presence of a CSO. Hence, the lack of separate spectral index information for both components prevents us from ruling out or confirming the binary black hole scenario for the double VLBA components.

\subsection{Comparison: Feasibility of the BBH model}
\begin{figure*}
    \centering
    \includegraphics[width=\linewidth]{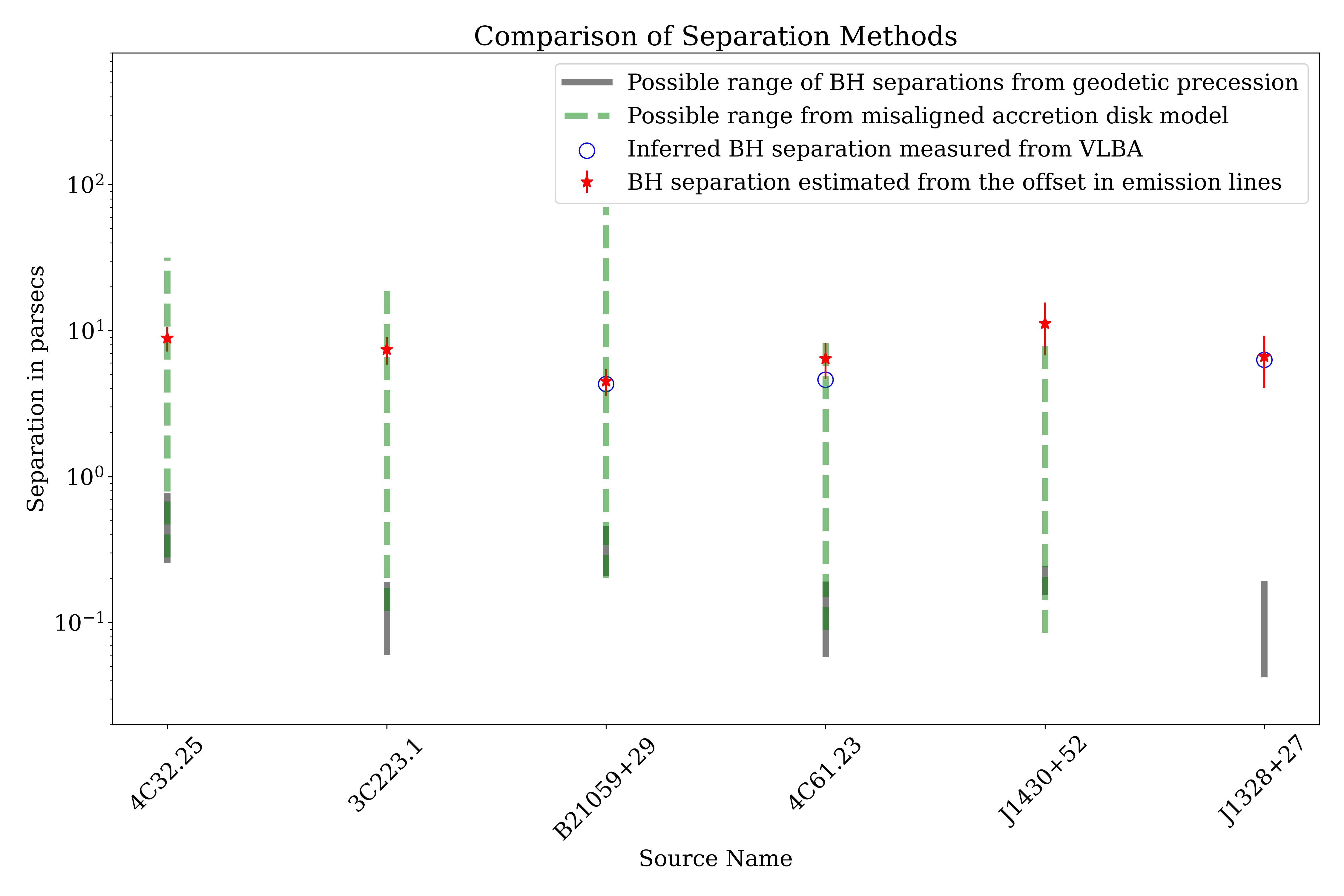} 
    \caption{Comparison of black hole separation methods. The plot illustrates the possible range of black hole separations from geodetic precession, misaligned accretion disk models estimated separation from the offset in emission lines, and inferred separation measured from VLBA. The separations derived from geodetic precession are inconsistent with those derived from the offsets in the emission lines and the VLBA compact components separation. However, the range of black hole separation predicted by the misaligned accretion disk model is consistent with the separation derived from emission lines. The correlation between the expected separation derived from the emission lines offset and the VLBA components' physical separation is significant.  It is important to note that any minor differences observed between the separations estimated from both methods could also, in principle, be introduced by projection effects.}
    \label{fig:comparison}
\end{figure*}

We have three indicators that predict the presence of BBH in three of our sample galaxies and at least two in the rest of the sample. A more conclusive test is to compare the physical properties of the BBH derived from these three separate diagnostics. Figure \ref{fig:comparison} shows a comparison between the BH separation predicted from the three different BBH diagnostics (see Sections~\ref{subsec:dpagnbbh}, \ref{sec:precession} and \ref{sec:dualcompactvlba}). 
The values predicted by the geodetic precession model are inconsistent with the other black hole separations and require the black holes to be closer than those predicted by the VLBA by at least one order of magnitude (see section~\ref{sec:bhestimates}). Note that all the methods result in black hole separations on parsec scales.

We estimate the range of black hole separations ($d_{BH}$) on parsec scales predicted by the misaligned accretion disk model for each target is obtained by inverting equation~\ref{PprecSMBBHS}.  Specifically, the lower boundary arises when $s=-2$ and $R_{out,p}=0.01$~pc, while the upper limit is for $s=0$ and $R_{out,p}$ equal to the tidal radius of the BBH. 
Although geodetic precession models disagree with the separations predicted by the other two methods, the misaligned accretion disk model can successfully explain the observed black hole separations. However, this can be partly attributed to the more significant number of free parameters in the model and uncertainties concerning our understanding of physical properties like the accretion disk radius.

On the other hand, the agreement between the predicted offset from the emission lines and the VLBA is highly encouraging. However, we note that projection effects affect both the separation estimates, and therefore, we should not expect a correlation with dispersions smaller than an order of magnitude. 

Finally, we compare the occurrence rate of X-shaped sources in DPAGN with extended radio morphology with that in a general sample of extended radio sources to uncover any potential correlation between the two signatures. 
Suppose we recovered a higher fraction of X/S-shaped sources from a sample of DPAGN with extended radio emission than a general sample of extended radio sources. In that case, it might suggest a common origin or, in other words, provide support for the binary black hole scenario. \citet[][the CC07 sample]{Cheung2007} discovered $\sim$100 X-shaped sources from a parent sample of $\sim$8000 extended FIRST sources with flux densities $>5$~mJy through visual inspection. The original FIRST source database at the time of the publication consisted of $\sim$800,000 sources. We used a similar approach, although with slight differences.

Our crossmatch of the DPAGN sample with FIRST returned $\sim$600 sources, out of which we found 30 extended sources and four X-shaped sources. Hence, we find that $\sim$13\% (four out of thirty) of DPAGN sources with extended radio emission show X-shaped morphology.
Since \cite{Cheung2007} does not clarify whether the sample of the X-shaped sources they identified is complete, we can only conclude that the typical fraction of cross-symmetric radio sources in a general sample of extended radio sources is $>$\,1.25\%. Several more cross-symmetric sources are reported in \cite{Yang2019} and \cite{Bera2020} owing to differences in classification approaches. The total number of X/S/Z shaped sources among the extended sources identified from the FIRST survey is only $\sim$400 to date ($\sim$5\%). Hence,  extended radio sources with DPAGN are at least twice as likely to possess an X-shaped radio morphology compared to a general sample of extended radio sources.
This conclusion remains valid whether we start from a sample of all or merely extended radio sources. More specifically, the fraction of cross-symmetric sources identified from the entire FIRST survey catalog (not only the extended sources) to date is $\sim$0.04\% as opposed to 0.7\% from a sample of FIRST sources that exhibit DPAGN.

While a higher probability for DPAGN to host more X-shaped sources argues for a BBH model in these systems, we would like to point out a few caveats. These probabilities are derived from a small sample and are prone to errors due to the small number statistics. Secondly, there might be several more X-shaped sources in the FIRST survey that remain to be discovered. Hence, while these numbers indicate the presence of underlying BBHs in many of these systems, a more rigorous statistical analysis is warranted to make more substantial claims.

Similarly, the high fraction of sources with dual radio peaks implies that our selection criterion is potentially useful in uncovering binary black holes for follow-up observations.




\section{Summary \& Conclusions}
\label{sec:summ}
We present new VLBA images at 5 and 15 GHz of six X-shaped radio galaxies with double-peaked emission lines in their optical spectra. New uGMRT images at Bands 3 and 5 (400~MHz and 1250~MHz) of three sources are also presented. We summarise below our primary findings. 

\begin{enumerate}
\item Three out of six sources exhibit dual compact radio components with at least one frequency using the VLBA. These could be dual cores, core-jet components, or CSOs. 
    
\item From the velocity separation of the blueshift and redshift line components detected in the optical spectra of these radio galaxies and their respective total black hole mass, we estimated the separation between the primary and secondary black holes under the hypothesis that the double lines are due to BBH in their nuclei, obtaining values roughly between 4 and 12 pc. These values are comparable to the compact component separations for the three sources with VLBA dual component detections.

\item The overall kpc-scale morphology of the radio galaxies in our sample is compatible with a jet/counter-jet precession scenario, even though some fine structural details in the radio images of some sources (e.g., systematic bending of the SE jet of 4C+32.25) are not recovered by our kinematic-only jet precession model. Extra mechanisms (e.g., ram-pressure phenomenon) must act upon the kiloparsec-scale jets in these cases. Our precession models suggest that the six sources have mildly relativistic jets ($\beta<0.5$), with precession cone axes oriented between $6\degr$ and $70\degr$ to the line of sight and semi-aperture precession angles between $3\degr$ and $49\degr$. Inferred jet precession periods on the observer's reference frame range from 1.5 to 500 Myr.

\item We have explored three different physical mechanisms behind jet precession: geodetic precession and precession induced by a secondary black hole in a non-coplanar orbit around the primary accretion disc, both possibilities implying the existence of a BBH system in the nuclei of our six sources, as well as the Bardeen-Petterson effect. Two of our six sources could not be analyzed in the context of the Bardeen-Petterson effect due to the lack of estimates of their bolometric luminosities. While geodetic precession predicts too long precession periods compared with the estimates derived from our precession modeling, the other two scenarios provide compatible precession periods for all the sources. It supports the jet/counter-jet precession scenario for the radio galaxies of our sample, even though these analyses do not univocally favor the BBH hypothesis.
 
\end{enumerate}

\section*{Acknowledgements}
We are grateful to the anonymous reviewer for their insightful comments, which helped improve our article.
B.S, C.O., and S.B. acknowledge support from
the Natural Sciences and Engineering Research Council
(NSERC) of Canada. P.K. acknowledges the support of the Department of Atomic Energy, Government of India, under the project 12-R\&D-TFR-5.02-0700. The National Radio Astronomy Observatory is a facility of the National Science Foundation operated under a cooperative agreement by Associated Universities, Inc.
Funding for the SDSS and SDSS-II has been provided by the Alfred P. Sloan Foundation, the Participating Institutions, the National Science Foundation, the U.S. Department of Energy, the National Aeronautics and Space Administration, the Japanese Monbukagakusho, the Max Planck Society, and the Higher Education Funding Council for England. The SDSS Web Site is http://www.sdss.org/.

\section{Data Availability Statement}
The data underlying this article will be shared on reasonable request to the corresponding author.

\bibliographystyle{mnras}

\bsp 
\label{lastpage}
\end{document}